\documentclass[10pt,a4paper,reqno]{amsart}
\usepackage{acronym}
\usepackage{amsfonts}
\usepackage{amsmath}
\usepackage{amssymb}
\usepackage{amsthm}
\usepackage{bm}
\usepackage{braket}
\usepackage{cite}
\usepackage{color}
\usepackage{geometry}
\usepackage{graphicx}
\usepackage{hyperref}
\usepackage{mathrsfs}
\usepackage{mathtools}
\usepackage{setspace}
\usepackage{stmaryrd}
\usepackage{subcaption}
\usepackage{tikz}
\usepackage{CJKutf8}

\usepackage{epsfig}
\usepackage{setspace}
\usepackage{booktabs}
\usepackage{threeparttable}
\usepackage{diagbox}

\newgeometry{top=2cm,bottom=2cm,outer=1.5cm,inner=1.5cm}
\newcommand{\bse}{\begin{subequations}}
\newcommand{\ese}{\end{subequations}}

\newtheorem{theorem}{Theorem}%[section]

\newtheorem{remark}[theorem]{Remark}

\numberwithin{equation}{section}

\title[Oceanic internal solitary wave interactions via the KP equation in a three-layer fluid with shear flow]{Oceanic internal solitary wave interactions via the KP equation in a three-layer fluid with shear flow}
\author{Junchao Sun}
\address[JS]{School of Mathematical Sciences, Ministry of Education Key Laboratory of Mathematics and Engineering Applications $\&$ Shanghai Key Laboratory of PMMP \\
East China Normal University \\ Shanghai 200241 \\ People's Republic of China}
\author{Xiaoyan Tang$^*$}
\address[XT]{School of Mathematical Sciences, Ministry of Education Key Laboratory of Mathematics and Engineering Applications $\&$ Shanghai Key Laboratory of PMMP \\
East China Normal University \\ Shanghai 200241 \\ People's Republic of China}
\author{Yong Chen}
\address[CY]{School of Mathematical Sciences, Ministry of Education Key Laboratory of Mathematics and Engineering Applications $\&$ Shanghai Key Laboratory of PMMP \\
East China Normal University \\ Shanghai 200241 \\ People's Republic of China}
\address[CY]{College of Mathematics and Systems Science \\ Shandong University of Science and Technology \\ Qingdao 266590 \\ People's Republic of China}
\email{xytang@sist.ecnu.edu.cn}

\begin{document}

\begin{abstract}
The various patterns of internal solitary wave interactions are complex phenomena in the ocean, susceptible to the influence of shear flow and density distributions. Satellite imagery serves as an effective tool for investigating these interactions, but usually does not provide information on the structure of internal waves and their associated dynamics. Considering a three-layer configuration that approximates ocean stratification, we analytically investigate two-dimensional internal solitary waves (ISW) in a three-layer fluid with shear flow and continuous density distribution using the (2+1)-dimensional Kadomtsev-Petviashvili (KP) model. Firstly, the KP equation is derived from the basic governing equations which include mass and momentum conservations, along with free surface boundary conditions. The coefficients of the KP equation are determined by the vertical distribution of fluid density, shear flow, and layer depth. Secondly, it is found that the interactions of ISW can be carefully classified into five types: ordinary interactions including O-type, asymmetric interactions including P-type, TP-type and TO-type, and Miles resonance. The genuine existence of these interaction types is observed from satellite images in the Andaman Sea, the Malacca Strait, and the coast of Washington state. Finally, the ``bright" and ``dark" internal solitary interactions are discovered in the three-layer fluid, which together constitute the fluctuating forms of oceanic ISW. It is revealed that shear flow is the primary factor to determine whether these types of interactions are ``bright" or ``dark". Besides, a detailed analysis is conducted to show how the ratio of densities influences the properties of these interactions, such as amplitude, angle, and wave width.
\end{abstract}

\maketitle

\section{Introduction}
Internal waves commonly occur in stratified fluids (oceans, lakes and fjords, etc) \cite{10, 7, 11}.
The first discoverer of internal waves was Nansen, whose vessel encountered the phenomenon of ``dead water" \cite{12} in
Arctic waters between $1893$ and $1896$, caused by the drag increasing effect of internal waves. In
the actual ocean environment, the density of seawater is  stable and continuously stratified, and any disturbance may
excite internal waves, making them ubiquitous in the ocean.
Internal solitary waves (ISW) are the most common and widely studied type of internal wave phenomena. Research has shown that in stratified seawater, ISW can be generated through the interaction of shear flow, internal tides, and wave-flow interactions in frontal regions \cite{13, 14}. ISW have a significant impact on the safety of offshore structures, the distribution of nutrients in water, and the propagation of acoustic waves, among other aspects. Therefore, in-depth research on ISW holds paramount theoretical and practical significance.

It is difficult to solve analytically the original equations for oceanic ISW, which may contain multiple dynamical processes at different scales and may even distort or obscure the features of our primary interest.
As the KdV equation is an important integrable equation applied in many physical fields, most
early investigations on ISW depended on it\cite{15, g}. Later, many low-dimensional equations such as the higher-order KdV equation, the Gardner equation, and the KdV-like equations
have been used to describe ISW \cite{16,17,18,19}. However, when considering the actual situation, we cannot ignore the three-dimensional spacial effect, so it is necessary to study ISW based on high dimensional models. Kadomtsev and Petviashvili derived a two-dimensional version of the KdV equation, known as the KP equation \cite{20}.
%Compared with the KdV equation, the KP equation has a richer set of solutions that are extremely important in physical applications. For instance, Zakharov and Shabat provided multiple soliton solutions of the KP equation \cite{21}, and Bordag et al. obtained the localized rational solutions of the KP equation termed as lumps \cite{22}.
Kataoka et al. earlier used the KP equation as a model for  ISW \cite{23}, and thereafter much work on ISW has focused on the KP equation and some other high-dimensional equations \cite{25,26,27}.
However, previous models have mostly employed a two-layer structure, where the density in each layer is treated as constant, and the influence of shear flow has not been considered. As we mentioned above, in the actual ocean environment, the continuous distribution of density and the presence of shear flow both play significant roles. Therefore, in the model development, it is essential for us to comprehensively consider these crucial factors.
To our knowledge, the KP equation has not been applied to ISW in a three-layer fluid. It is found that the vertical stratification has a clearly pronounced three-layer structure in the ocean \cite{45, three, w}. Therefore, it is also necessary to introduce a three-layer model
to explain the basic features of the internal wave field in such environments.

ISW often interact with each other during their propagations \cite{d, 5, 6},
and these interactions can threaten the safety of offshore structures, ships, and submarines. Many
theoretical analysis on the interactions of ISW have been carried out to help
people understand them further. For instance, Yu et al. studied the ordinary interactions based
on the KP type equation \cite{28}. Wang et al. described the Mach interactions observed in the Strait of Georgia \cite{29}. Yuan et al. simulated diffraction and oblique interactions \cite{31}.
However, focusing on only one type of interaction may not be enough to understand the interaction of ISW within the ocean.
A noteworthy study is the one conducted by Xue et al., who analyzed the interactions among three different types of internal waves in the Mid-Atlantic Bight based on satellite imagery \cite{30}. The theoretical foundation of Xue's work is established upon the ``bright" interaction solutions of the KP-type equation  with constant coefficients,
\begin{equation}\label{1-1}
\left(\eta_t+c_0 \eta_x+c_1 \eta \eta_x+c_2 \eta_{x x x}\right)_x+\frac{c_0}{2} \eta_{y y}=0 .
\end{equation}
The internal wave interactions are not only common in the mid-Atlantic but also in other marine regions.
These internal wave interactions, even though they occur within the ocean, exhibit a surface feature on the sea surface (manifesting as  a small modulation on the surface roughness) that can be captured by satellite imagery.
For instance, the ERS-2 satellite has collected a significant amount of internal wave data in the Andaman Sea, including interactions between ISW \cite{iw}. Furthermore, photographs taken by astronauts (STS036-082-76) showcase the intricate patterns formed when ISW collide in the southern African maritime region \cite{iw}. Nowadays, satellite images have become an  efficient tool to study internal wave interactions, however, satellite images by themselves usually do not provide information about the internal wave structure and its associated motions, which motivates us to study these interactions in depth.

It is worth mentioning that Kodama and Biondini have theoretically studied three fundamental types of 2-soliton interaction structures for the KP-II equation \cite{32,33,34}. On the other hand, Ablowitz and Baldwin observed  two types of interactions for shallow-water waves on flat beaches \cite{ab}, and mathematically described them using the KP-type equation

\begin{equation}\label{1-2}
\frac{\partial}{\partial x}\left(\frac{1}{\sqrt{g h}} \eta_t+\eta_x+\frac{3}{2 h} \eta \eta_x+\frac{h^2 \gamma}{2} \eta_{x x x}\right)+\frac{1}{2} \eta_{y y}=0.
\end{equation}
The aforementioned studies primarily focused on describing surface waves. Due to the typically challenging nature of observing fluctuations occurring within the internal environment of fluids, research on systems involving internal interaction types remains relatively limited. Our work aims to describe ISW in the ocean by establishing a reasonable model and to explore their internal interaction patterns by drawing on surface wave theory, as well as to validate the feasibility of the theoretical study through satellite images.

%\begin{figure}[!htbp]
%\centering
%{
%\label{fig:2} %% label for first subfigure
%\includegraphics[width=13cm]{2-IWI.png}}
%%\hspace{0.5in}
%\caption{The interaction of internal solitary waves (Astronaut photograph (STS036-082-76) acquired on 1 March 1990 at 1254 UTC), from http://www.internalwaveatlas.com/}
%%\label{fig:subfig} %% label for entire figure
%\end{figure}

The rest of the paper is organized as follows. In Section 2, a (2+1) dimensional KP model is derived  for describing oceanic ISW. In Section 3, the coefficients of the KP equation, as defined by the particular vertical distribution of fluid density, layer depth and properties of shear flow, are explicitly calculated and analysed in detail in a three-layer fluid.
In Section 4, the ``bright" and ``dark"  ISW are diagnosed, and the internal solitary wave interactions are carefully categorized into five types, which can reflect interaction patterns in the real ocean. It is revealed that shear flow is the primary factor determining the generation of ``bright" and ``dark" interactions, while the density ratio also influences properties such as amplitude, angle, and wave width in these interactions.

\section{Derivation of the  KP model for internal waves}
\subsection{Governing equations}
In order to derive the KP equation modeling  oceanic internal waves , we start from the inviscid, incompressible, and layered fluid. The basic governing equations, consisting of the mass and momentum conservation equations  in three-dimensions, are

\begin{align}\label{1}
&\rho\frac{d u}{d t}+\frac{\partial p}{\partial x}=0,\\ \label{2}
&\rho\frac{d v}{d t}+\frac{\partial p}{\partial y}=0,\\ \label{3}
&\rho\frac{d w}{d t}+\frac{\partial p}{\partial z}+\rho \text{\sl g}=0,\\ \label{4}
&\frac{d \rho}{d t}=0,\\ \label{5}
&\frac{\partial u}{\partial x}+\frac{\partial v}{\partial y}+\frac{\partial w}{\partial z}=0,
\end{align}
where $x$, $y$ and $z$ are the spatial coordinates, and $u$, $v$ and $w$ are the fluid velocities in the $x$, $y$ and $z$ directions, respectively, $\rho$ is the fluid density, $p$ is the pressure, and $\text{\sl g}$ is the gravitational acceleration. The material derivative $d/dt$ is expressed in the following form,

\begin{equation}\label{6}
\frac{d }{d t}=\frac{\partial }{\partial t}+u\frac{\partial }{\partial x}+v\frac{\partial }{\partial y}+w\frac{\partial }{\partial z}.
\end{equation}

Consider that the fluid takes the rigid boundary $z=-h$ as the lower boundary, the free surface $z=\psi(x,y,t)$ as the upper boundary, and the equilibrium position of the upper boundary is $z=0$.
Therefore, the boundary conditions of the governing equations are

%
%\begin{figure}[!htbp]
%\centering
%{
%\includegraphics[width=10cm,height=7cm]{3-FS(1).png}}
%%\hspace{0.5in}
%\caption{A schematic representation of the boundaries}
%\label{fig-1}
%\end{figure}

\begin{align}\label{7}
&w=0\mid_{z=-h},\\ \label{8}
&p=0\mid_{z=\psi(x,y,t)},\\ \label{9}
&w=\frac{\partial \psi}{\partial t}+u\frac{\partial \psi}{\partial x}+v\frac{\partial \psi}{\partial y}\mid_{z=\psi(x,y,t)},
\end{align}
where $\psi(x,y,t)$  is the vertical displacement of the free surface. Eq. \eqref{9} ensures that the vertical velocity at the free surface coincides with the vertical velocity inside the fluid.

Introducing the characteristic length $h_{0}$, the characteristic density $\overline{\rho}$, and the characteristic buoyancy frequency  $N_{0}=\text{\sl g}\Delta\overline{\rho}/h_{0}\overline{\rho}$, we can define the dimensionless variables  as follows,

\begin{align}\label{10}
&(x,~y,~z,~t)=(h_{0}\widetilde{x},~h_{0}\widetilde{y},~h_{0}\widetilde{z},~\frac{1}{N_{0}}\widetilde{t}),\\ \label{11}
&(u,~v,~w)=(h_{0}N_{0}\widetilde{u},~h_{0}N_{0}\widetilde{v},~h_{0}N_{0}\widetilde{w}),\\ \label{12}
&(h,~\varphi,~\rho,~p,)=(h_{0}\widetilde{h},~h_{0}\widetilde{\varphi},~\overline{\rho}\widetilde{\rho},~\overline{\rho}h_{0}\text{\sl g}\widetilde{\rho}),
\end{align}
then by substituting Eqs. \eqref{10}-\eqref{12} into Eqs. \eqref{1}-\eqref{5} and the boundary conditions \eqref{7}-\eqref{8}, and ignoring the superscripts of dimensionless variables, the governing equations and boundary conditions in the dimensionless form can be obtained,
\begin{align} \label{13}
&\rho\frac{d u}{d t}+\frac{1}{\sigma}\frac{\partial p}{\partial x}=0,\\ \label{14}
&\rho\frac{d v}{d t}+\frac{1}{\sigma}\frac{\partial p}{\partial y}=0,\\ \label{15}
&\rho\frac{d w}{d t}+\frac{1}{\sigma}\left(\frac{\partial p}{\partial z}+\rho\right)=0,\\ \label{16}
&\frac{d \rho}{d t}=0,\\ \label{17}
&\frac{\partial u}{\partial x}+\frac{\partial v}{\partial y}+\frac{\partial w}{\partial z}=0,\\ \label{18}
&w=0\mid_{z=-h},\\ \label{19}
&p=0\mid_{z=\psi(x,~y,~t)},\\ \label{20}
&w=\frac{\partial \psi}{\partial t}+u\frac{\partial \psi}{\partial x}+v\frac{\partial \psi}{\partial y}\mid_{z=\psi(x,~y,~t)},
\end{align}
where $\sigma=h_{0}N_{0}^{2}/\text{\sl g}$ is small in the ocean conditions.

\subsection{Semi-Lagrangian form}

In this subsection, we further transform the governing equations and boundary conditions into the semi-Lagrangian form.
A new variable $\zeta(x,~y,~z,~t)$ is introduced to represent the vertical displacement of a fluid particle from its rest position, which is obviously related to $w$ as
\begin{equation} \label{21}
w=\frac{d \zeta}{d t}.
\end{equation}

According to Eq. \eqref{15}, the pressure $p$ is denoted by
\begin{equation} \label{22}
p(x,y,z,t)=-\int_{0}^{z} \rho_{0}(z')dz'+\sigma q(x,y,z,t),
\end{equation}
where the function $q(x,y,z,t)$ is the complex integral function.

Suppose the density of the fluid is $\rho_{0}(z)$ in the rest state. Therefore, the density of the perturbed fluid reads $\rho(x,y,z,t)=\rho_{0}(z-\zeta(x,~y,~z,~t))$. Hence, the Lagrangian coordinate is introduced as
\begin{equation}\label{23}
k=z-\zeta(x,y,z,t),
\end{equation}
and the density of the fluid in the perturbed state is reformulated as
\begin{equation} \label{24}
\rho(x,y,z,t)=\rho_{0}(k).^{}
\end{equation}

%\begin{equation}
%w=\frac{d \zeta}{d t}=\frac{\partial \zeta}{\partial t}+u\frac{\partial \zeta}{\partial x}+v\frac{\partial \zeta}{\partial y}+w\frac{\partial \zeta}{\partial z},
%\end{equation}
Based on Eqs. \eqref{23} and \eqref{24}, we obtain

\begin{equation}
\begin{aligned}\label{25}
\frac{d \rho}{d t}&=\frac{d \rho_{0}}{d t}=\frac{\partial \rho_{0}}{\partial t}+u\frac{\partial \rho_{0}}{\partial x}+v\frac{\partial \rho_{0}}{\partial y}+w\frac{\partial \rho_{0}}{\partial z}\\
%&=-\frac{\partial \rho_{0}}{\partial k}\frac{\partial \zeta}{\partial t}-u\frac{\partial \rho_{0}}{\partial k}\frac{\partial \zeta}{\partial x}-v\frac{\partial \rho_{0}}{\partial k}\frac{\partial \zeta}{\partial y}+w\frac{\partial \rho_{0}}{\partial k}(1-\frac{\partial \zeta}{\partial z})\\
&=\frac{\partial \rho_{0}}{\partial k}\left(w-\frac{\partial \zeta}{\partial t}-u\frac{\partial \zeta}{\partial x}-v\frac{\partial \zeta}{\partial y}-w\frac{\partial \zeta}{\partial z}\right)\\
&=\frac{\partial \rho_{0}}{\partial k}\left(w-\frac{d \zeta}{d t}\right)=0.
\end{aligned}
\end{equation}

It is obvious that the introduction of Lagrangian coordinates makes Eq. \eqref{16} identically satisfied. Now let us
derive the partial derivatives of an arbitrary function $f(x,y,z,t)$ in Eulerian coordinates with respect to time and space, as well as the form of its material derivative.
Denoting $f(x,y,z,t)=f'(x,y,k,t)$, we have

\begin{equation}
\begin{aligned}\label{26}
&\frac{\partial f}{\partial t}=\frac{\partial f'}{\partial t}-\frac{\partial f'}{\partial k} \frac{\partial \zeta}{\partial t},\\
&\frac{\partial f}{\partial x}=\frac{\partial f'}{\partial x}-\frac{\partial f'}{\partial k} \frac{\partial \zeta}{\partial x},\\
&\frac{\partial f}{\partial y}=\frac{\partial f'}{\partial y}-\frac{\partial f'}{\partial k} \frac{\partial \zeta}{\partial y},\\
&\frac{\partial f}{\partial z}=\frac{\partial f'}{\partial k}-\frac{\partial f'}{\partial k} \frac{\partial \zeta}{\partial z},
\end{aligned}
\end{equation}
which lead to the material derivative as
\begin{equation}
\begin{aligned}\label{27}
\frac{d f}{d t}=\frac{\partial f'}{\partial t}+u \frac{\partial f'}{\partial x}+v \frac{\partial f'}{\partial y}.
\end{aligned}
\end{equation}

Letting $\zeta(x, y, z, t)=\eta(x, y, k, t)$, the partial derivatives of $\zeta(x, y, z, t)$ in Eq. \eqref{26} can be determined as

\begin{equation}
\begin{aligned}\label{28}
\zeta_z=\frac{\eta_k}{1+\eta_k}, \quad \zeta_x=\frac{\eta_x}{1+\eta_k}, \quad \zeta_y=\frac{\eta_y}{1+\eta_k},\quad \zeta_t=\frac{\eta_t}{1+\eta_k}.
\end{aligned}
\end{equation}

Using Eqs. \eqref{26}-\eqref{28} and ignoring the superscripts of the functions, we rewrite Eqs. \eqref{13}-\eqref{15} and \eqref{17} in the new coordinates as

\begin{align}\label{29}
&\rho_0(k)\left(\frac{\partial u}{\partial t}+u \frac{\partial u}{\partial x}+v\frac{\partial u}{\partial y}\right)+\frac{\partial q}{\partial x}-\frac{1}{1+\frac{\partial \eta}{\partial k}} \frac{\partial q}{\partial k} \frac{\partial \eta}{\partial x}=0,\\ \label{30}
&\rho_0(k)\left(\frac{\partial v}{\partial t}+u \frac{\partial v}{\partial x}+v\frac{\partial v}{\partial y}\right)+\frac{\partial q}{\partial y}-\frac{1}{1+\frac{\partial \eta}{\partial k}} \frac{\partial q}{\partial k} \frac{\partial \eta}{\partial y}=0,\\ \label{31}
&\rho_0(k)\left(\frac{\partial w}{\partial t}+u \frac{\partial w}{\partial x}+v\frac{\partial w}{\partial y}\right)+\frac{1}{1+\frac{\partial \eta}{\partial k}} \frac{\partial q}{\partial k}+\frac{1}{\sigma}\left[\rho_0(k)-\rho_0(k+\eta)\right]=0,\\ \label{32}
&\frac{\partial u}{\partial x}+\frac{\partial v}{\partial y}+\frac{\partial w}{\partial k}-\frac{1}{1+\frac{\partial \eta}{\partial k}}\left(\frac{\partial u}{\partial k} \frac{\partial \eta}{\partial x}+\frac{\partial v}{\partial k} \frac{\partial \eta}{\partial y}+\frac{\partial w}{\partial k} \frac{\partial \eta}{\partial k}\right)=0.
\end{align}

%\begin{equation}
%\xi=\varepsilon(x-\lambda t),~ \eta=\varepsilon^{2}y,~ \zeta=\varepsilon^{2}z,~ \tau=\varepsilon^{3}t,\\
%\end{equation}\\

From Eqs. \eqref{21} and \eqref{28}, we get

\begin{equation}
\begin{aligned} \label{33}
&w=\frac{\partial \eta}{\partial t}+u \frac{\partial \eta}{\partial x}+v \frac{\partial \eta}{\partial y}.\\
\end{aligned}
\end{equation}

It is important to mention that since Eq. \eqref{21} satisfies the boundary condition $\zeta=\psi$ at $z=\psi$, Eq. \eqref{33} is still satisfied on the boundary.
Under the new coordinates, the boundary conditions become
\begin{align}\label{34}
&\left.\int_{0}^{\eta}\rho_0\left(k^{\prime}\right) d k^{\prime}=\sigma q\right|_{k=0},\\ \label{35}
&\left.\eta=0\right|_{k=-h}.
\end{align}

Finally, Eqs. \eqref{29}-\eqref{32} are reduced to the following three equations by using Eq. \eqref{33} and eliminating the function $q(x,y,z,t)$,

\begin{equation}
\begin{aligned}\label{36}
&\frac{\partial}{\partial k}\left\{\rho_0(k)\left(\frac{\partial u}{\partial t}+u \frac{\partial u}{\partial x}+v\frac{\partial u}{\partial y}\right)\right\}-\rho_0(k) N^2(k) \frac{\partial \eta}{\partial x} \\
&-\left(1+\frac{\partial \eta}{\partial k}\right)\frac{\partial}{\partial x}\left\{\rho_0(k)\left(\frac{\partial}{\partial t}+u \frac{\partial}{\partial x}+v \frac{\partial}{\partial y}\right)^2 \eta\right\}\\
&+ \frac{\partial \eta}{\partial x}\frac{\partial}{\partial k}\left\{\rho_0(k)\left(\frac{\partial}{\partial t}+u \frac{\partial}{\partial x}+v \frac{\partial}{\partial y}\right)^2 \eta\right\}=0,
\end{aligned}
\end{equation}

\begin{equation}
\begin{aligned}\label{37}
&\frac{\partial}{\partial k}\left\{\rho_0(k)\left(\frac{\partial v}{\partial t}+u \frac{\partial v}{\partial x}+v\frac{\partial v}{\partial y}\right)\right\}-\rho_0(k) N^2(k) \frac{\partial \eta}{\partial y} \\
&-\left(1+\frac{\partial \eta}{\partial k}\right)\frac{\partial}{\partial y}\left\{\rho_0(k)\left(\frac{\partial}{\partial t}+u \frac{\partial}{\partial x}+v \frac{\partial}{\partial y}\right)^2 \eta\right\}\\
&+ \frac{\partial \eta}{\partial y}\frac{\partial}{\partial k}\left\{\rho_0(k)\left(\frac{\partial}{\partial t}+u \frac{\partial}{\partial x}+v \frac{\partial}{\partial y}\right)^2 \eta\right\}=0,
\end{aligned}
\end{equation}

\begin{equation}
\begin{aligned}\label{38}
\left(1+\frac{\partial\eta}{\partial k}\right)\left(\frac{\partial u}{\partial x}+\frac{\partial v}{\partial y}\right)+\frac{\partial^{2}\eta}{\partial t\partial k}+u\frac{\partial^{2}\eta}{\partial x\partial k}+v\frac{\partial^{2}\eta}{\partial y\partial k}=0,
\end{aligned}
\end{equation}
where
\begin{equation}
\begin{aligned}\label{39}
N^2(k)=-\frac{1}{\sigma \rho_0(k)} \frac{d \rho_0(k)}{d k}.
\end{aligned}
\end{equation}

The boundary conditions \eqref{34} and \eqref{35}, after eliminating the function $q(x,y,z,t)$, are
\begin{align}\label{40}
&\frac{\partial \eta}{\partial x}=-\sigma\left(\frac{\partial u}{\partial t}+u \frac{\partial u}{\partial x}+v\frac{\partial u}{\partial y}\right)-\left.\sigma \frac{\partial \eta}{\partial x}\left(\frac{\partial}{\partial t}+u \frac{\partial}{\partial x}+v \frac{\partial}{\partial y}\right)^2 \eta\right|_{k=0},\\ \label{41}
&\eta=\left.0\right|_{k=-h}.
\end{align}

It is noted that in the semi-Lagrangian form, the original governing equations are reformed as Eqs. \eqref{36}-\eqref{38}  with the boundary conditions \eqref{40} and \eqref{41}. In this way, the number of the equations and the boundary conditions are both reduced. However, this semi-Lagrangian method leads to an increase in the order of nonlinearity to the fourth order, whereas the original controlling model has only two orders of nonlinearity.

\begin{figure}[!htbp]
\centering
{
\includegraphics[width=10cm,height=7cm]{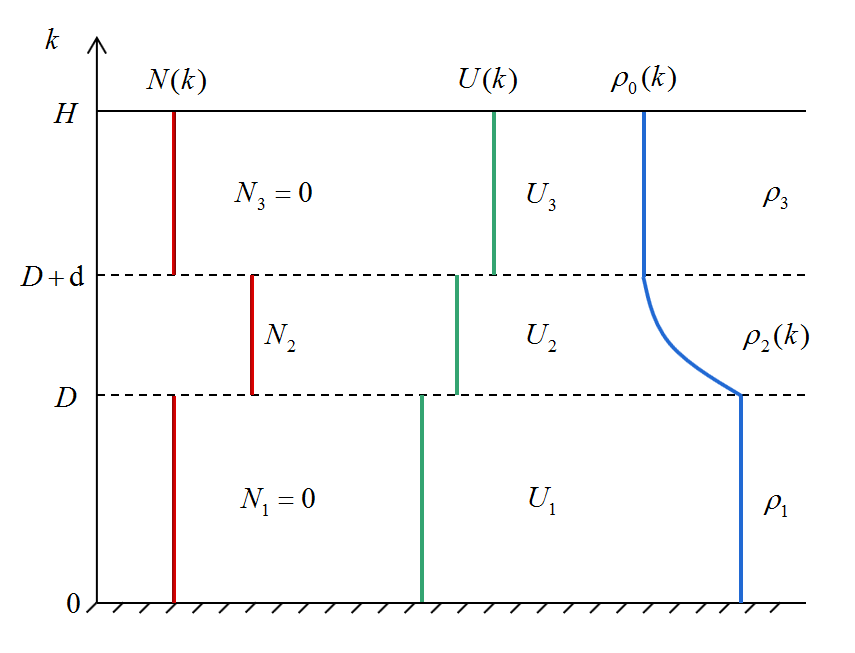}}
%\hspace{0.5in}
\caption{A schematic representation of the three-layer system.}
\label{fig-2}
\end{figure}

\subsection{Derivation of the KP equation}

It is remarkable that Eqs. \eqref{36}-\eqref{38} are still a complicated set of nonlinear equations, so it is  not easy to obtain  explicit general solutions. Here, we utilize the  multiple scale method\cite{16} to derive  the KP equation modeling two-dimensional ISW.

For the discussion of nonlinear long waves, the coordinate extension method in the long wave approximation, namely, the Gardner-Morikawa transform, can be used. Introduce a small parameter $\epsilon$ to have the slow variables

\begin{equation}
\begin{aligned}\label{42}
&X=\epsilon x,~Y=\epsilon y,~T=\epsilon t,
\end{aligned}
\end{equation}
%According to Eq.(42), we have
which gives
\begin{equation}
\begin{aligned}\label{43}
&\frac{\partial}{\partial x}=\epsilon\frac{\partial}{\partial X},~\frac{\partial}{\partial y}=\epsilon \frac{\partial}{\partial Y},~\frac{\partial}{\partial t}=\epsilon \frac{\partial}{\partial T} .
\end{aligned}
\end{equation}

Separate the velocity field in the $x$-direction into an elementary component $U(k)$ and a perturbation $u^{\prime}(x, y, k, t)$, while the velocity field in the $y$-direction only has a perturbation component $v^{\prime}(x, y, k, t)$, i.e.,
\begin{align}\label{44}
&u(x, y, k, t)=U(k)+u^{\prime}(x, y, k, t),\\ \label{45}
&v(x, y, k, t)=v^{\prime}(x, y, k, t).
\end{align}
%Indeed, even if we decompose the velocity field $v(x, y, k, t)$ in the y-direction into a fundamental component and a perturbation, we still get this fundamental component equal to zero in the subsequent derivation.

Based on Eqs. \eqref{43}-\eqref{45}, Eqs. \eqref{36}-\eqref{38} can be rewritten as
\begin{equation}
\begin{aligned}\label{46}
&\frac{\partial}{\partial k}\left\{\rho_0(k)\left[\frac{\partial u^{\prime}}{\partial T}+\left(U(k)+u^{\prime}\right) \frac{\partial u^{\prime}}{\partial X}+v^{\prime}\frac{\partial u^{\prime}}{\partial Y}\right]\right\}-\rho_0(k) N^2(k) \frac{\partial \eta}{\partial X}\\
&-\epsilon^2 \left(1+\frac{\partial \eta}{\partial k}\right)\frac{\partial}{\partial X}\left\{\rho_0(k)\left[\frac{\partial}{\partial T}
+\left(U(k)+u^{\prime}\right) \frac{\partial}{\partial X}+v^{\prime}\frac{\partial }{\partial Y}\right]^2 \eta\right\} \\
&+\epsilon^2  \frac{\partial \eta}{\partial X}\frac{\partial}{\partial k}\left\{\rho_0(k)\left[\frac{\partial}{\partial T}
+\left(U(k)+u^{\prime}\right) \frac{\partial}{\partial X}+v^{\prime}\frac{\partial }{\partial Y}\right]^2 \eta\right\} =0,\\
\end{aligned}
\end{equation}

\begin{equation}
\begin{aligned} \label{47}
&\frac{\partial}{\partial k}\left\{\rho_0(k)\left[\frac{\partial v^{\prime}}{\partial T}+\left(U(k)+u^{\prime}\right)\frac{\partial v^{\prime}}{\partial X}+v^{\prime}\frac{\partial v^{\prime}}{\partial Y}\right]\right\}-\rho_0(k) N^2(k) \frac{\partial \eta}{\partial Y}\\
&-\epsilon^2 \left(1+\frac{\partial \eta}{\partial k}\right)\frac{\partial}{\partial Y}\left\{\rho_0(k)\left[\frac{\partial}{\partial T}
+\left(U(k)+u^{\prime}\right) \frac{\partial}{\partial X}+v^{\prime}\frac{\partial }{\partial Y}\right]^2 \eta\right\} \\
&+\epsilon^2  \frac{\partial \eta}{\partial Y}\frac{\partial}{\partial k}\left\{\rho_0(k)\left[\frac{\partial}{\partial T}
+\left(U(k)+u^{\prime}\right) \frac{\partial}{\partial X}+v^{\prime}\frac{\partial }{\partial Y}\right]^2 \eta\right\} =0,\\
\end{aligned}
\end{equation}

\begin{equation}
\begin{aligned} \label{48}
\left(1+\frac{\partial\eta}{\partial k}\right)\left(\frac{\partial u^{\prime}}{\partial X}+\frac{\partial v^{\prime}}{\partial Y}\right)+\frac{\partial^2 \eta}{\partial T \partial k}+\left(U(k)+u^{\prime}\right) \frac{\partial^2 \eta}{\partial X \partial k}+v^{\prime} \frac{\partial^2 \eta}{\partial Y \partial k}=0,
\end{aligned}
\end{equation}

with the boundary conditions

\begin{align}\label{49}
&\frac{\partial \eta}{\partial X}=-\sigma\left(\frac{\partial u^{\prime}}{\partial T}+\left(U(k)+u^{\prime}\right) \frac{\partial u^{\prime}}{\partial X}+v^{\prime}\frac{\partial u^{\prime}}{\partial Y}\right)\\ \nonumber
&\left.-\epsilon^2 \sigma \frac{\partial \eta}{\partial X}\left(\frac{\partial}{\partial T}+\left(U(k)+u^{\prime}\right) \frac{\partial}{\partial X}+v^{\prime}\frac{\partial}{\partial Y}\right)^2 \eta\right|_{k=0},\\ \label{50}
&\eta=\left.0\right|_{k=-h}.
\end{align}

Then, introduce new variables
\begin{equation}
\begin{aligned} \label{51}
\xi=X-c T, \quad \theta=\epsilon Y, \quad \tau=\mu T,
\end{aligned}
\end{equation}
where $c$ is the velocity of the  long wave, and $\mu=\epsilon^{2}$. Consequently, we have
\begin{equation}
\begin{aligned} \label{52}
\frac{\partial}{\partial T} =-c \frac{\partial}{\partial \xi}+\mu \frac{\partial}{\partial \tau},~~\frac{\partial}{\partial X} =\frac{\partial}{\partial \xi},~~
\frac{\partial}{\partial Y} =\epsilon\frac{\partial}{\partial \theta}.\\
\end{aligned}
\end{equation}

\begin{figure}[!htbp]
  \begin{center}
     \includegraphics[width=15cm]{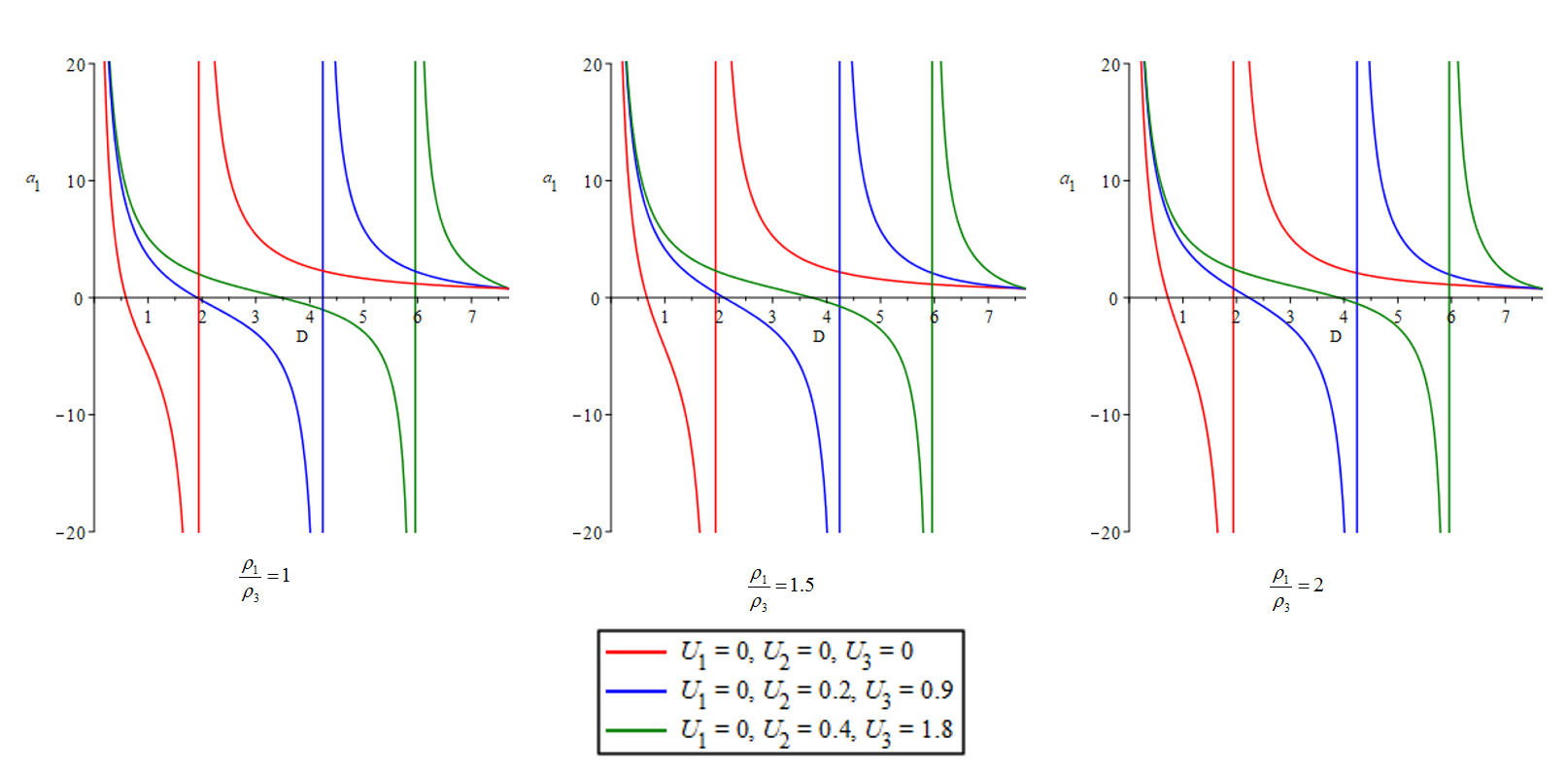}
     \caption{Coefficient $a_{1}$  with $H=8$, $c=4$ and $d=0.3$.}
       \label{fig-3}
       \end{center}
\end{figure}

Substituting Eq. \eqref{52} into Eqs. \eqref{46}-\eqref{49} arrives at

\begin{equation}
\begin{aligned}\label{53}
\frac{\partial}{\partial k}\left\{\rho_0(k)(U(k)-c) \frac{\partial u^{\prime}}{\partial \xi}\right\}-\rho_0(k) N^2(k) \frac{\partial \eta}{\partial \xi}=F,
\end{aligned}
\end{equation}

\begin{equation}
\begin{aligned}\label{54}
\frac{\partial}{\partial k}\left\{\rho_0(k)(U(k)-c) \frac{\partial v^{\prime}}{\partial \xi}\right\}-\epsilon\rho_0(k) N^2(k) \frac{\partial \eta}{\partial \theta}=F^{\prime},
\end{aligned}
\end{equation}

\begin{equation}
\begin{aligned}\label{55}
\frac{\partial u^{\prime}}{\partial \xi}+(U(k)-c) \frac{\partial^2 \eta}{\partial \xi \partial k}=G,
\end{aligned}
\end{equation}

\begin{equation}
\begin{aligned}\label{56}
\frac{\partial \eta}{\partial \xi} &+\sigma(U(k)-c) \frac{\partial u^{\prime}}{\partial \xi}=-\left.\sigma\left(\mu \frac{\partial u^{\prime}}{\partial \tau}+u^{\prime} \frac{\partial u^{\prime}}{\partial \xi}+\epsilon v^{\prime}\frac{\partial v^{\prime}}{\partial \theta}+\mu \frac{\partial \eta}{\partial \xi} H\right)\right|_{k=0},
\end{aligned}
\end{equation}
where
\begin{equation}
\begin{aligned}\label{57}
F=&-\frac{\partial}{\partial k}\left[\rho_0(k)\left(\mu \frac{\partial u^{\prime}}{\partial \tau}+u^{\prime} \frac{\partial u^{\prime}}{\partial \xi}+\epsilon v^{\prime} \frac{\partial u^{\prime}}{\partial \theta}\right)\right] \\
&+\mu\left(1+\frac{\partial \eta}{\partial k}\right) \frac{\partial}{\partial \xi}\left(\rho_0 H\right)-\mu \frac{\partial \eta}{\partial \xi} \frac{\partial}{\partial k}\left(\rho_0 H\right), \\
F'=&-\frac{\partial}{\partial k}\left[\rho_0(k)\left(\mu \frac{\partial v^{\prime}}{\partial \tau}+u^{\prime} \frac{\partial v^{\prime}}{\partial \xi}+\epsilon v^{\prime} \frac{\partial v^{\prime}}{\partial \theta}\right)\right] \\
&+\epsilon\mu\left(1+\frac{\partial \eta}{\partial k}\right) \frac{\partial}{\partial \theta}\left(\rho_0 H\right)-\epsilon\mu \frac{\partial \eta}{\partial \theta} \frac{\partial}{\partial k}\left(\rho_0 H\right), \\
G=&-\mu \frac{\partial^2 \eta}{\partial \tau \partial k }-\frac{\partial}{\partial \xi}\left(u^{\prime} \frac{\partial \eta}{\partial k}\right)-\epsilon \frac{\partial v^{\prime}}{\partial \theta}-\epsilon \frac{\partial }{\partial \theta}(v^{\prime}\frac{\partial\eta}{\partial k}),\\
H=&\left[(U(k)-c) \frac{\partial}{\partial \xi}+\mu \frac{\partial}{\partial \tau}\right.\left.+u^{\prime} \frac{\partial}{\partial \xi}+\epsilon v^{\prime} \frac{\partial}{\partial \theta}\right]^2 \eta.
\end{aligned}
\end{equation}

\begin{figure}[!htbp]
  \begin{center}
     \includegraphics[width=15cm]{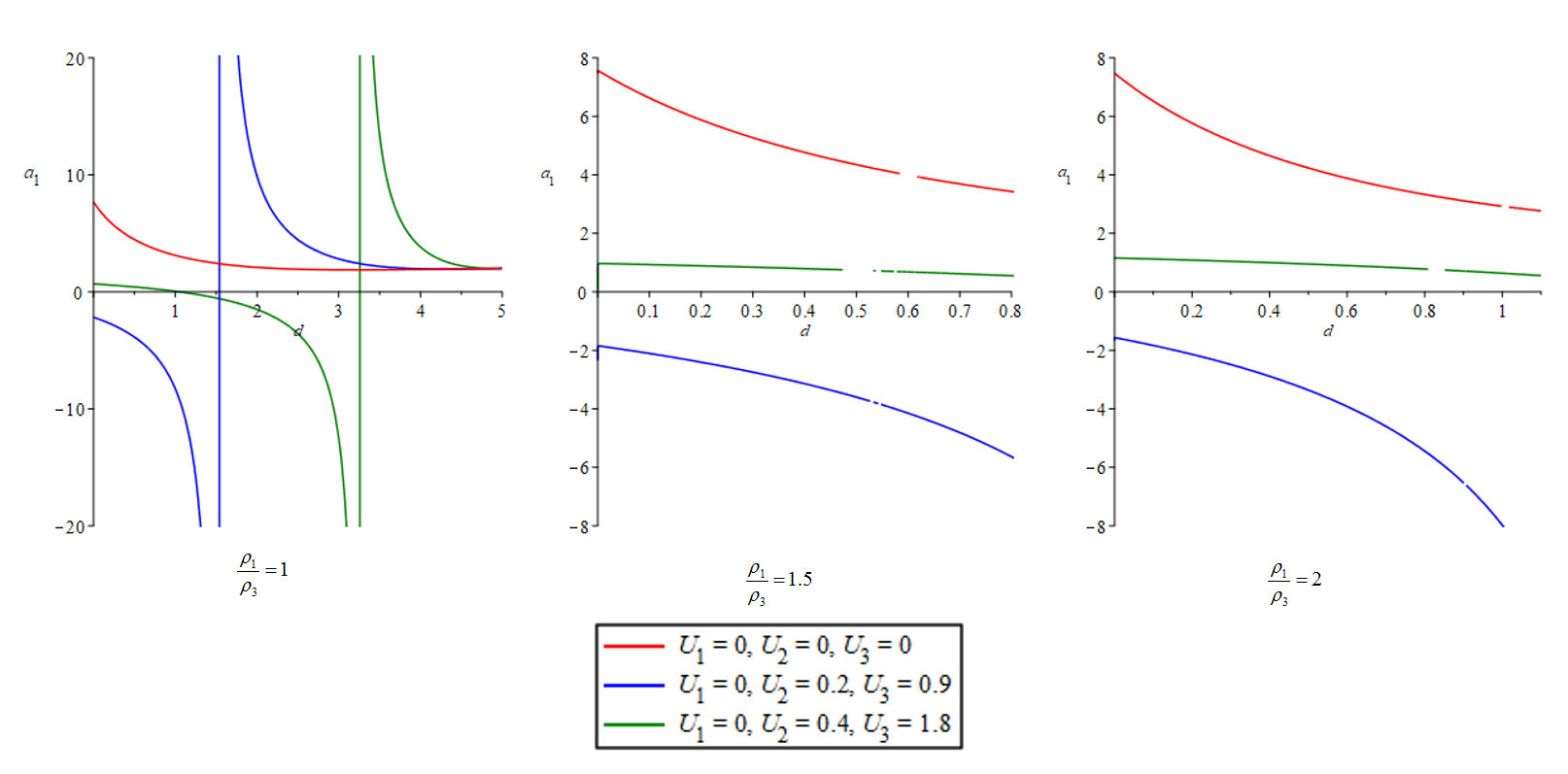}
     \caption{Coefficient $a_{1}$  with $H=8$, $c=4$ and $D=3$.}
       \label{fig-4}
       \end{center}
\end{figure}

By eliminating  $u^{\prime}$ on the left-hand side of the above equations, Eqs. \eqref{53}  and \eqref{55} can be further simplified and degenerated to one equation,

\begin{equation}
\begin{aligned}\label{58}
\frac{\partial}{\partial k}\left\{\rho_0(k)(U(k)-c)^2 \frac{\partial^2 \eta}{\partial \xi \partial k}\right\}+\rho_0(k) N(k)^2 \frac{\partial \eta}{\partial \xi}=M,
\end{aligned}
\end{equation}
with the boundary condition
\begin{equation}
\begin{aligned}\label{60}
\frac{\partial \eta}{\partial \xi}=\sigma(U(k)-c)^2 \frac{\partial^2 \eta}{\partial \xi \partial k}-\sigma(U(k)-c) G+\left.\sigma H_1\right|_{k=0},
\end{aligned}
\end{equation}
where
\begin{align}\label{59}
&M=\frac{\partial}{\partial k}\left\{\rho_0(k)(U(k)-c) G\right\}-F,\\ \label{61}
&H_1=-\left(\mu \frac{\partial u^{\prime}}{\partial \tau}+u^{\prime} \frac{\partial u^{\prime}}{\partial \xi}+\epsilon v^{\prime} \frac{\partial v^{\prime}}{\partial \theta}+\mu \frac{\partial \eta}{\partial \xi} H\right).
\end{align}

Expanding  $\eta(\xi,\theta, k,\tau)$, $u^{\prime}(\xi,\theta, k,\tau)$ and $v^{\prime}(\xi,\theta, k,\tau)$ in the following asymptotic form,
\begin{equation}
\begin{aligned}\label{62}
\begin{aligned}
&\eta(\xi,\theta, k,\tau)=\mu A(\xi,\theta,\tau) \Phi(k)+\mu^2 \eta_1(\xi,\theta, k,\tau)+\mu^3 \eta_2(\xi,\theta, k,\tau)+\ldots ~,\\
&u^{\prime}(\xi,\theta, k,\tau)=\mu u_0(\xi,\theta, k,\tau)+\mu^2 u_1(\xi,\theta, k,\tau)+\mu^3 u_2(\xi,\theta, k,\tau)+\ldots~,\\
&v^{\prime}(\xi,\theta, k,\tau)=\epsilon^{3} v_1(\xi,\theta, k,\tau)+\epsilon^{5} v_2(\xi,\theta, k,\tau)+\epsilon^{7} u_2(\xi,\theta, k,\tau)+\ldots~,
\end{aligned}
\end{aligned}
\end{equation}
substituting them into Eqs. \eqref{54}, \eqref{55}, \eqref{58} and the boundary conditions \eqref{50} and \eqref{60}, and then collecting the terms of the same order in $\epsilon$,  we obtain the perturbation problems at each order.

At the order of $\mu$ and $\epsilon\mu$, we have

\begin{equation}\label{63}
\mu:
\left\{
\begin{aligned}
&\frac{d}{d k}\left[\rho_0(k)(U(k)-c)^2 \frac{d \Phi}{d k}\right]+\rho_0 N^2 \Phi=0, \\
&\Phi=\left.0\right|_{k=-h}, \\
&\Phi=\left.\sigma(U(k)-c)^2 \frac{d \Phi}{d k}\right|_{k=0},
\end{aligned}
\right.
\end{equation}

\begin{equation}\label{64}
\mu:
\begin{aligned}
\frac{\partial u_{0}}{\partial \xi}+(U(k)-c)\frac{d \Phi}{d k}\frac{\partial A}{\partial \xi}=0.
\end{aligned}
\end{equation}
%where Eq. (63) is the eigenvalue problem and Eq. (64) gives the relationship between $u_{0}$ and $A$.

\begin{equation}\label{65}
\epsilon\mu:
\begin{aligned}
\frac{\partial }{\partial k}\left[\rho_0(k)(U(k)-c)\frac{\partial v_{1}}{\partial \xi}\right]-\rho_0(k)N^{2}(k)\Phi\frac{\partial A}{\partial \theta}=0.
\end{aligned}
\end{equation}

%\begin{equation}
%L \frac{\partial \eta}{\partial \xi}=M
%\end{equation}
%
%Here $L$ is the linear operator,
%\begin{equation}
%L=\frac{\partial}{\partial k}\left[\rho_0(k)(U(k)-c)^2 \frac{\partial}{\partial k}\right]+\rho_0 N^2
%\end{equation}

%Then we express the solvability conditions (compatibility conditions) of Eq. (58) and boundary conditions (50), (60) as an equation without boundary terms.

Rewriting Eq. \eqref{58}  as
\begin{equation}
\begin{aligned} \label{66}
M \Phi=\frac{\partial }{\partial k}\left[\rho_0(k)(U(k)-c)^2\Phi\frac{\partial^{2} \eta}{\partial \xi\partial k}\right]-\frac{\partial }{\partial k}\left[\rho_0(k)(U(k)-c)^2\frac{\partial \Phi}{\partial k}\frac{\partial \eta}{\partial \xi}\right],
\end{aligned}
\end{equation}
and the integrating it  with the boundary conditions \eqref{50} and \eqref{60} results in

\begin{equation}
\begin{aligned}\label{67}
\int_{-h}^0 M \Phi d k=\sigma\left\{\rho_0(k)(U(k)-c)^2 \frac{d \Phi}{d k}\left[(U(k)-c) G-H_1 \right] \right\} _{k=0},
\end{aligned}
\end{equation}
%This compatibility condition has a boundary term.
which can be reformulated via Eq. \eqref{59} as

\begin{equation}
\begin{aligned} \label{68}
&\int_{-h}^0 \frac{\partial}{\partial k}\left\{\rho_0(k)(U(k)-c) G\Phi\right\}d k-\int_{-h}^0 F\Phi dk -\int_{-h}^0\rho_0(k)(U(k)-c) G \frac{\partial \Phi}{\partial k} dk\\
&=\sigma \left [ \rho_0(k)(U(k)-c)^2 \frac{d \Phi}{d k}H_{1}\right ] _{k=0}+\sigma \left [ \rho_0(k)(U(k)-c)^2 \frac{d \Phi}{d k}(U(k)-c) G\right ] _{k=0}.\\
\end{aligned}
\end{equation}

\begin{figure}[!htb]
  \begin{center}
     \includegraphics[width=15cm]{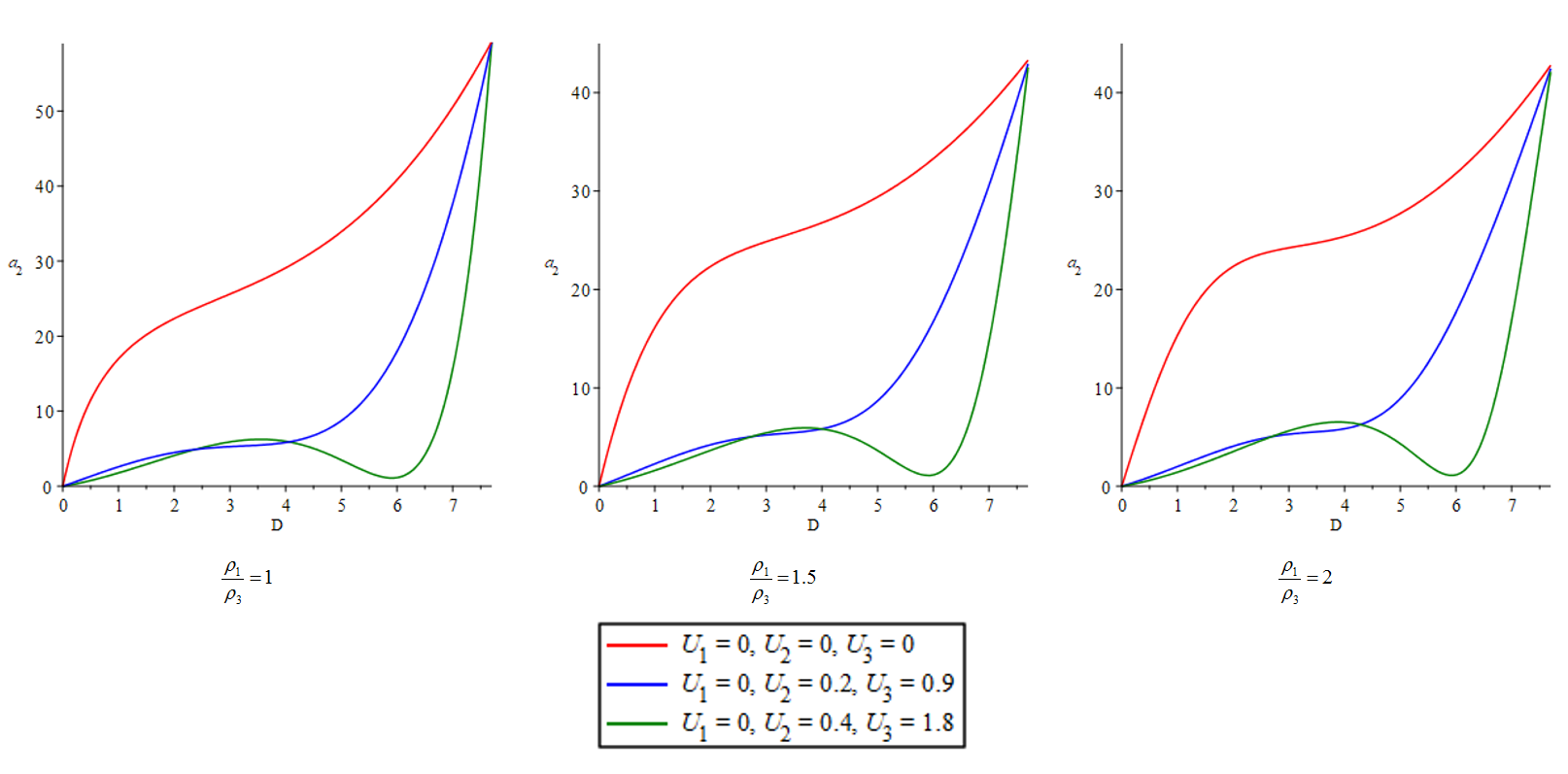}
     \caption{Coefficient $a_{2}$  with $H=8$, $c=4$ and $d=0.3$.}
\label{fig-5}
       \end{center}
\end{figure}

According to Eq. \eqref{63}, we obtain
\begin{equation}
\begin{aligned} \label{69}
&\int_{-h}^0 F\Phi dk +\int_{-h}^0\rho_0(k)(U(k)-c) G \frac{\partial \Phi}{\partial k} dk-\sigma \left [ \rho_0(k)(U(k)-c)^2 \frac{d \Phi}{d k}H_{1}\right ] _{k=0}=0.
\end{aligned}
\end{equation}

Based on Eq. \eqref{57}, we easily get

\begin{equation}
\begin{aligned} \label{70}
F=&-\frac{\partial}{\partial k}\left[\rho_0(k)\left(\mu \frac{\partial u^{\prime}}{\partial \tau}+u^{\prime} \frac{\partial u^{\prime}}{\partial \xi}+\epsilon v^{\prime} \frac{\partial u^{\prime}}{\partial \theta}+ \mu\frac{\partial \eta}{\partial \xi}H          \right)\right] \\
&+\mu\left(1+\frac{\partial \eta}{\partial k}\right) \frac{\partial}{\partial \xi}\left(\rho_0 H\right)+\mu \rho_0 H\frac{\partial }{\partial k}\left( \frac{\partial \eta}{\partial \xi} \right),
\end{aligned}
\end{equation}
which can be expressed as
\begin{equation} \label{71}
F=\frac{\partial F_{1}}{\partial k}+\frac{\partial F_{2}}{\partial \xi},
\end{equation}
with
\begin{equation} \label{72}
F_{1}=\rho_0 H_{1},~~F_{2}=\mu\rho_0 H \left(1+\frac{\partial \eta}{\partial k}\right)  .
\end{equation}

It is necessary to note that
\begin{equation} \label{73}
\int_{-h}^0 \frac{\partial }{\partial k}(F_{1}\Phi) dk= \left [ \sigma \rho_0(k)(U(k)-c)^2 \frac{d \Phi}{d k}H_{1}\right ] _{k=0}.
\end{equation}

Hereafter, the substitution of Eqs. \eqref{71} and \eqref{73} into Eq. \eqref{69} leads to

\begin{equation}
\begin{aligned} \label{74}
&\int_{-h}^0 \frac{\partial F_{2}}{\partial \xi}\Phi dk -\int_{-h}^0 F_{1}\frac{d \Phi}{d k}dk +\int_{-h}^0\rho_0(k)(U(k)-c) G \frac{d \Phi}{d k} dk=0,
\end{aligned}
\end{equation}
%and it does not contain boundary terms.
where the boundary terms are removed naturally.

From the expansions of Eqs. \eqref{54} and \eqref{74},  the  order of $\mu^{2}$ gives

\begin{equation} \label{75}
\begin{aligned}
&\int_{-h}^0 \frac{\partial }{\partial \xi}\left[\rho_0(k)(U(k)-c)^2 \frac{\partial^{2} }{\partial \xi^{2}}(A \Phi)   \right]\Phi dk+  \int_{-h}^0  \rho_0(k) \left( \frac{\partial u_{0}}{\partial \tau}+ u_{0}\frac{\partial u_{0}}{\partial \xi}\right) \frac{d \Phi}{d k} dk\\
&+\int_{-h}^0  \rho_0(k) (U(k)-c)\left[ -\frac{\partial A}{\partial \tau}\frac{d \Phi}{d k}  - \frac{\partial }{\partial \xi}\left( u_{0}A\frac{d \Phi}{d k} \right)-\frac{\partial v_{1}}{\partial \theta} \right] \frac{d \Phi}{d k} dk=0.
\end{aligned}
\end{equation}

\begin{figure}[!htbp]
  \begin{center}
     \includegraphics[width=15cm]{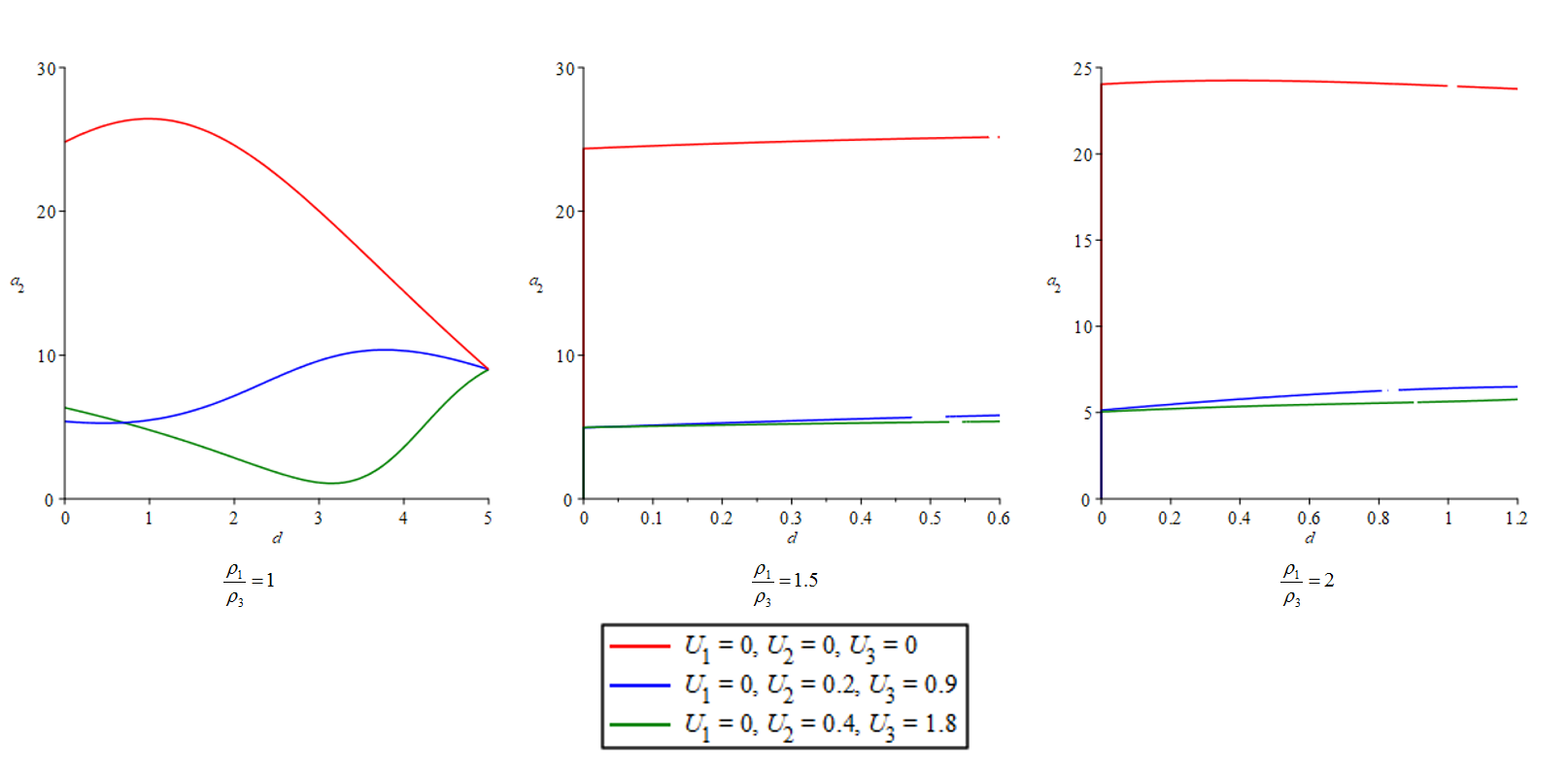}
     \caption{Coefficient $a_{2}$  with $H=8$, $c=4$ and $D=3$.}
  \label{fig-6}
       \end{center}
\end{figure}

Finally, substituting Eqs. \eqref{64} and \eqref{65} into Eq. \eqref{75}, we arrive at the KP equation
\begin{equation}\label{76}
\frac{\partial }{\partial \xi}\left(\frac{\partial A}{\partial \tau}+a_{1}A\frac{\partial A}{\partial \xi}+a_{2}\frac{\partial ^{3}A}{\partial \xi^{3}}\right)+a_{3}\frac{\partial ^{2}A}{\partial \theta^{2}}=0,\\
\end{equation}
where
\begin{equation}\label{77}
\begin{aligned}
&a_{1}=\frac{3\int_{-h}^0 \rho_0(k)(U(k)-c)^2 \left(\frac{d \Phi}{d k}\right)^{3} dk}{2\int_{-h}^0 \rho_0(k)(c-U(k))\left(\frac{d \Phi}{d k}\right)^{2} dk},\\
&a_{2}=\frac{\int_{-h}^0 \rho_0(k)(U(k)-c)^2 \Phi^{2} dk}{2\int_{-h}^0 \rho_0(k)(c-U(k))\left(\frac{d \Phi}{d k}\right)^{2} dk},\\
&a_{3}=\frac{-\int_{-h}^0 \int_{0}^k \rho_0(k')N(k')^2 \Phi dk' \frac{d \Phi}{d k}dk}{2\int_{-h}^0 \rho_0(k)(c-U(k))\left(\frac{d \Phi}{d  k}\right)^{2} dk}.\\
\end{aligned}
\end{equation}

%Eq. \eqref{76} is an integrable equation for which we can solve some exact solutions, before that we need to explain the exact expressions of the coefficients.

Here, the variable coefficients of Eq. \eqref{76}  are closely related to many physical quantities, giving them an advantage compared to Eqs. \eqref{1-1} and \eqref{1-2}. Besides, other  equations describing internal waves, such as the KdV equation and Boussinesq equation, usually deal with two layers of fluid with constant density of the upper and lower layers. However, when such a two-layer stratification is considered, the coefficient $a_{3}$ in Eq. \eqref{76} will be zero, and Eq. \eqref{76} will be reduced to the  KdV equation. In the next section, a three-layer fluid with continuous density distribution  is investigated in detail.

\begin{figure}[!htbp]
  \begin{center}
     \includegraphics[width=15cm]{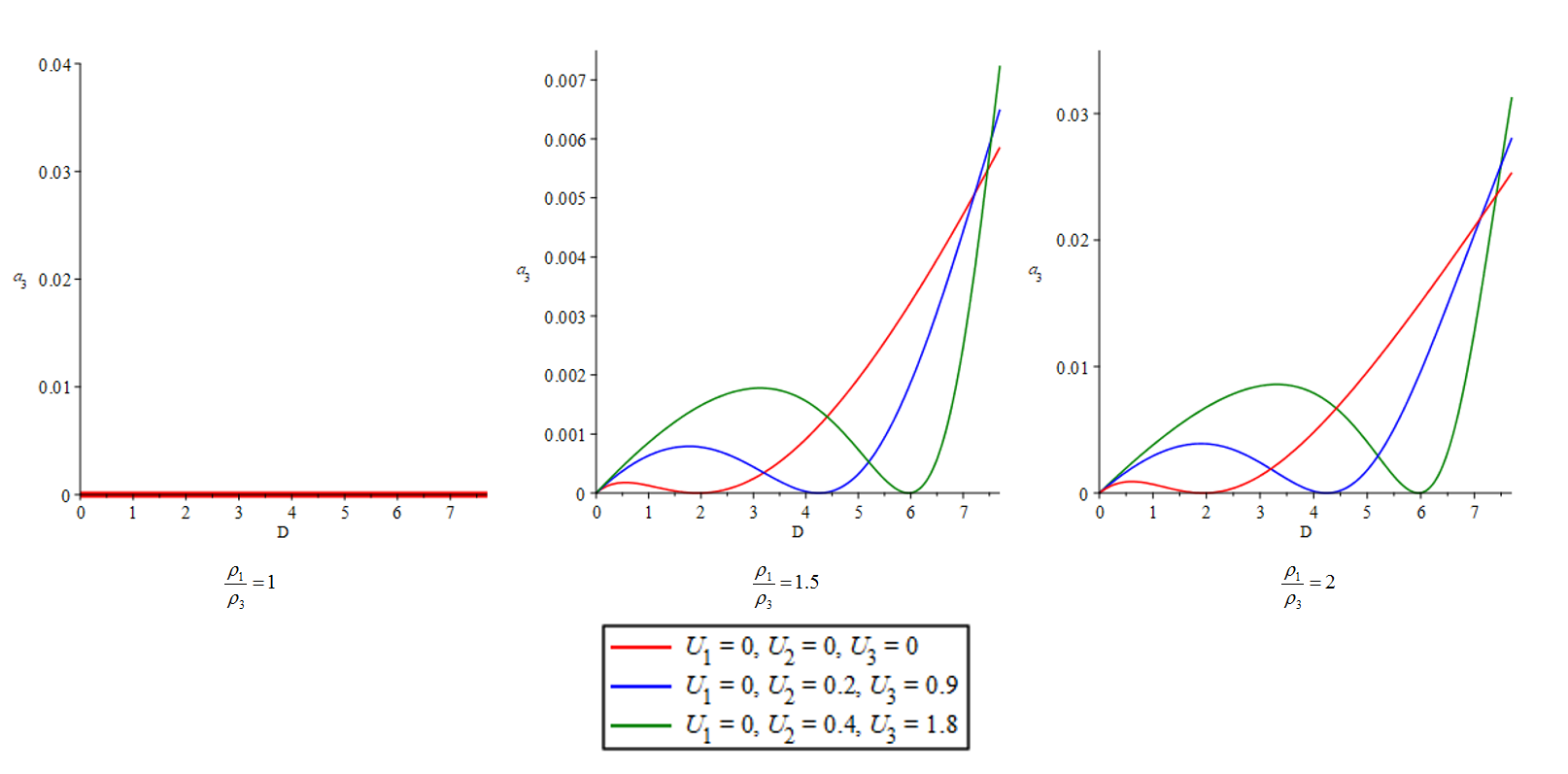}
     \caption{Coefficient $a_{3}$  with $H=8$, $c=4$ and $d=0.3$.}
  \label{fig-7} %% label for first subfigure
       \end{center}
\end{figure}

\section{Coefficients of the KP equation for a three-layer fluid}

The three-layer structure of fluid  is displayed in Fig. \ref{fig-2}. The depths of the upper, middle and lower layers are $H-d-D$, $d$ and $D$, respectively. Densities $\rho_{3}$ and $\rho_{1}$ of  the upper and lower layers are constant, while, the density of the middle layer is a depth-dependent function $\rho_{2}(k)$, and $\rho_{1}>\rho_{2}(k)>\rho_{3}$. Such a stratified structure is also similar to the stratification found in the ocean.
Based on the density distribution, the buoyancy frequencies of the upper and lower layers are zero, i.e., $N_{3}$=$N_{1}$=0, and the middle layer has a constant buoyancy frequency $N_{2}$.

Each fluid layer is assumed to have a constant current velocity $U_{i}$($i=1, 2, 3$), with $U_{1}<U_{2}<U_{3}$, In this situation, the shear flow in the three-layer fluid  is supposed to be a piecewise constant function, and this shear flow is affected by the Kelvin-Helmholtz instability, which can be neglected when considering long waves, and the fluid we are investigating can be regarded as an effective approximation to a system with a continuous shear flow.

\subsection{Calculation of the coefficients}
As mentioned above, the specific formulas for the density become

\begin{equation}\label{78}
\rho_{0}(k):
\left\{
\begin{aligned}
&\rho_{1},~~~~~~ ~~\quad \quad\quad\quad\quad\quad\quad\quad\quad\quad\quad0 \leq k<D,\\
&\rho_{2}(k)=\rho_{3}e^{\frac{1}{d}\ln\frac{\rho_{1}}{\rho_{3}}(D+d)}e^{-\frac{1}{d}\ln\frac{\rho_{1}}{\rho_{3}}k},~~~D \leq k\leq D+d,\\
&\rho_{3},~~~~~~ ~~\quad \quad\quad\quad\quad\quad\quad\quad\quad\quad\quad D+d <k\leq H.
\end{aligned}
\right.
\end{equation}

The corresponding buoyancy frequencies are

\begin{equation}\label{79}
N(k):
\left\{
\begin{aligned}
&N_{1}=0,~~~~~~~~~~~\quad\quad\quad0 \leq k<D,\\
&N_{2}=\sqrt{\frac{1}{\sigma d} \ln \frac{\rho_1}{\rho_3}},~~~~~\quad D \leq k\leq D+d,\\
&N_{3}=0,~~~~~~~~~~~\quad\quad\quad D+d <k\leq H.
\end{aligned}
\right.
\end{equation}

The modal function $\Phi$ is obtained from Eqs. \eqref{53}, \eqref{79} and the eigenvalue problem \eqref{63},

\begin{equation}\label{80}
\Phi=
\left\{
\begin{aligned}
&\frac{1}{D}k,~~~~~~~~~~~~\quad\quad\quad\quad\quad\quad\quad\quad\quad\quad\quad\quad\quad\quad\quad\quad\quad\quad\quad\quad ~~0\leq k<D,\\
&\frac{-e^{\frac{(k-D-b)p_{1}+q_{1}}{2d\sigma(-U_{2}+c)}}+e^{\frac{(-k+D+b)p_{1}+q_{1}}{2d\sigma(-U_{2}+c)}}-e^{\frac{(-k+D)p_{1}+q_{2}}{2d\sigma(-U_{2}+c)}}+e^{\frac{(k-D)p_{1}+q_{2}}{2d\sigma(-U_{2}+c)}}}{e^{\frac{dp_{1}+q_{3}}{2d\sigma(-U_{2}+c)}}-e^{-\frac{dp_{1}+q_{3}}{2d\sigma(-U_{2}+c)}}},~ D \leq k\leq D+d,\\
&\frac{k+\sigma(U_{3}-C)^{2}-H}{\sigma(U_{3}-C)^{2}-H+D+d},~~~~~~~~~~~~~~~~~~\quad\quad\quad\quad\quad\quad\quad\quad\quad D+d <k\leq H,
\end{aligned}
\right.
\end{equation}
where  the maximum values of the modal functions of the upper and lower layers have been set one. The expressions for $p_{1}$, $q_{1}$, $q_{2}$ and $q_{3}$ are determined as follows
\begin{equation}\label{81}
\begin{aligned}
&p_{1}=\sqrt{\sigma\left(\sigma(-U_{2}+c)^2 \ln \left(\frac{\rho_{1}}{\rho_{3}}\right)-4 d\right) \ln \left(\frac{\rho_{1}}{\rho_{3}}\right)},\\
&q_{1}=\sigma \ln \left(\frac{\rho_{1}}{\rho_{3}}\right)(d+D+k)(-U_{2}+c),\\
&q_{2}=\sigma \ln \left(\frac{\rho_{1}}{\rho_{3}}\right)(D+k)(-U_{2}+c),\\
&q_{3}=\sigma \ln \left(\frac{\rho_{1}}{\rho_{3}}\right)(d+2 D)(-U_{2}+c).\\
\end{aligned}
\end{equation}

Finally, Eqs. \eqref{78}-\eqref{81} are substituted into Eq. \eqref{77} to obtain the coefficients of the KP equation, which are presented  in Appendix A.

 \begin{figure}[!htbp]
  \begin{center}
     \includegraphics[width=15cm]{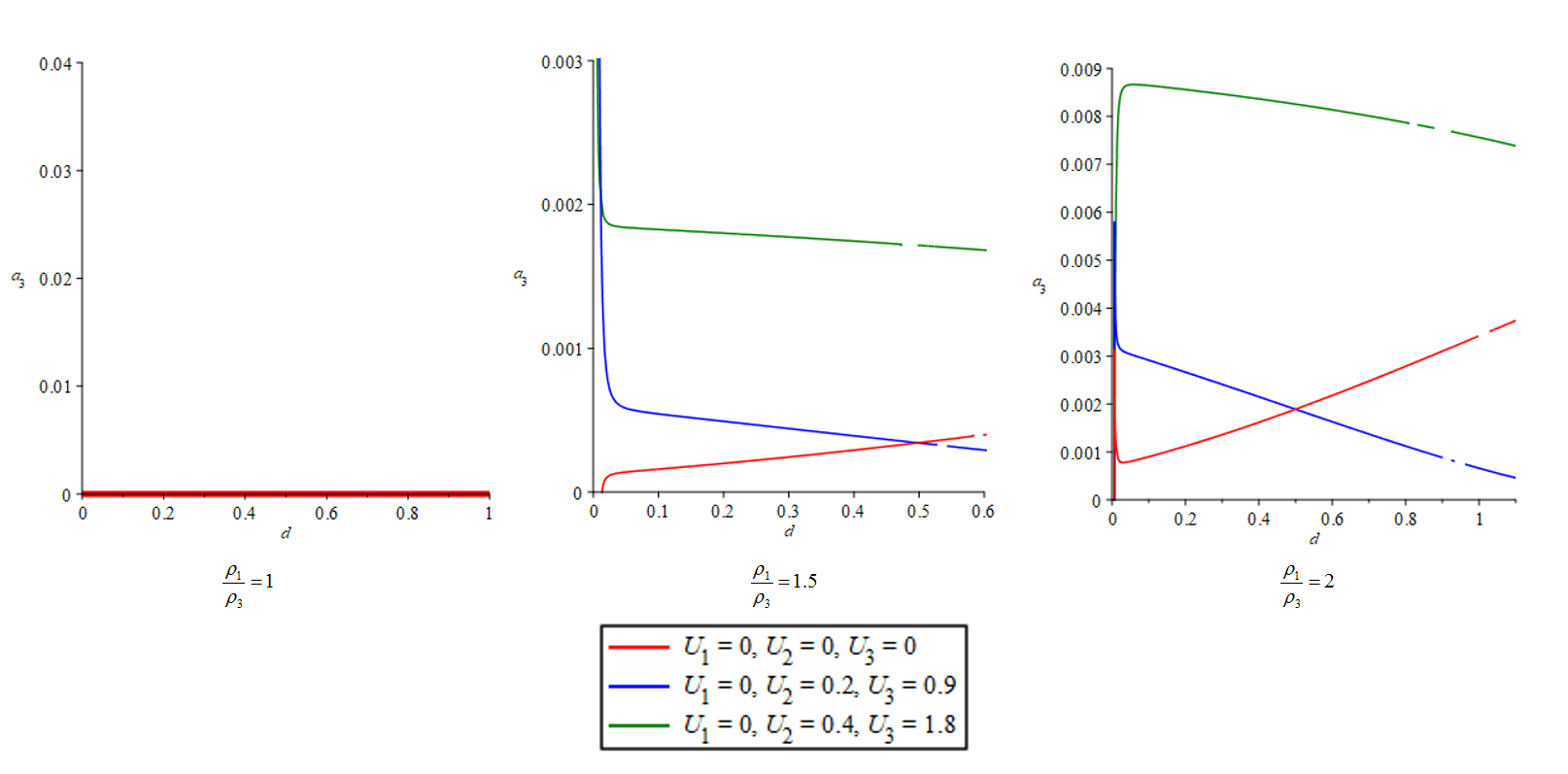}
     \caption{Coefficient $a_{3}$  with $H=8$, $c=4$ and $D=3$.}
   \label{fig-8}
       \end{center}
\end{figure}

\subsection{Analysis of the coefficients}
We show  graphics of the coefficients $a_{1}$, $a_{2}$ and $a_{3}$ as functions of  the shear flows, ratio of density and depth of the lower and middle layers in Figs. \ref{fig-3}-\ref{fig-8}. It should be noted that though $\rho_{1}/\rho_{3}>1$,  we also show the related figures with $\rho_{1}/\rho_{3}=1$ for a better comparison.

The value of  $a_{1}$ can be positive or negative, and  has an infinite value at $D = 0$, that is, the depth of the lower layer is zero, see  Fig. \ref{fig-3}. Besides, $a_{1}$ has singular and zero points about $D$, and their locations  can be understood as  where the nonlinear effects are very strong and very weak, respectively. It is found that increasing the value of $\rho_{1}/\rho_{3}$ would make  the zero points move in the increasing direction of  $D$, but has no impact on  the singular  points. However, as the shear flow increases, the zero and singular positions move in the positive direction along the $D$-axis.

It is revealed from Fig. \ref{fig-4}  that the variation of $\rho_{1}/\rho_{3}$ can affect the range of  $d$ except the case of $\rho_{1}/\rho_{3}=1$. As a matter of fact, in the case of $\rho_{1}/\rho_{3}>1$, the minimum value of $d$ cannot be $0$.  The maximum value of $d$ can only reach the position  where the curve starts to appear as a gap, and it is clear from the expression  $d<\frac{1}{4}\sigma(-U_{2}+c)^2\ln(\frac{\rho_{1}}{\rho_{3}})$.
Obviously, $d$ is a small value with respect to $H$ and $D$, which explains well why we usually consider the middle layer as a thin layer. Specifically, increasing the ratio of the densities $\rho_{1}$ and $\rho_{3}$ increases the range of  $d$. The presence of  shear flows also affects the value of $d$, but the effect is very weak compared to the change caused by densities. Likewise, these results are not found when $\rho_{1}/\rho_{3}=1$.

The dispersion coefficient $a_{2}$ can not be negative, and goes to zero at $D = 0$.  The ratio of the densities $\rho_{1}$ and $\rho_{3}$ has no significant effect on the basic trend of the curves about $a_{2}$. However, the presence of the shear flows greatly changes the trend of the curves, and this change is  more pronounced as the value of the shear flows increase, as shown in Fig. \ref{fig-5}.

The reason for this phenomenon can be manifested by comparing Figs. \ref{fig-3} and  \ref{fig-5}, where one can clearly observe that with the increase of the shear flows, the dispersion term $a_{2}$ has local maximum and minimum values, corresponding to the positions of zero and singular points in the nonlinear term $a_{1}$, respectively. It is  indicated that the dispersion effect becomes weaker at the locations where the nonlinear effect is suddenly enhanced (i.e., singularity locations) and vice versa.

\begin{figure}[!htb]
  \begin{center}
     \includegraphics[width=15cm]{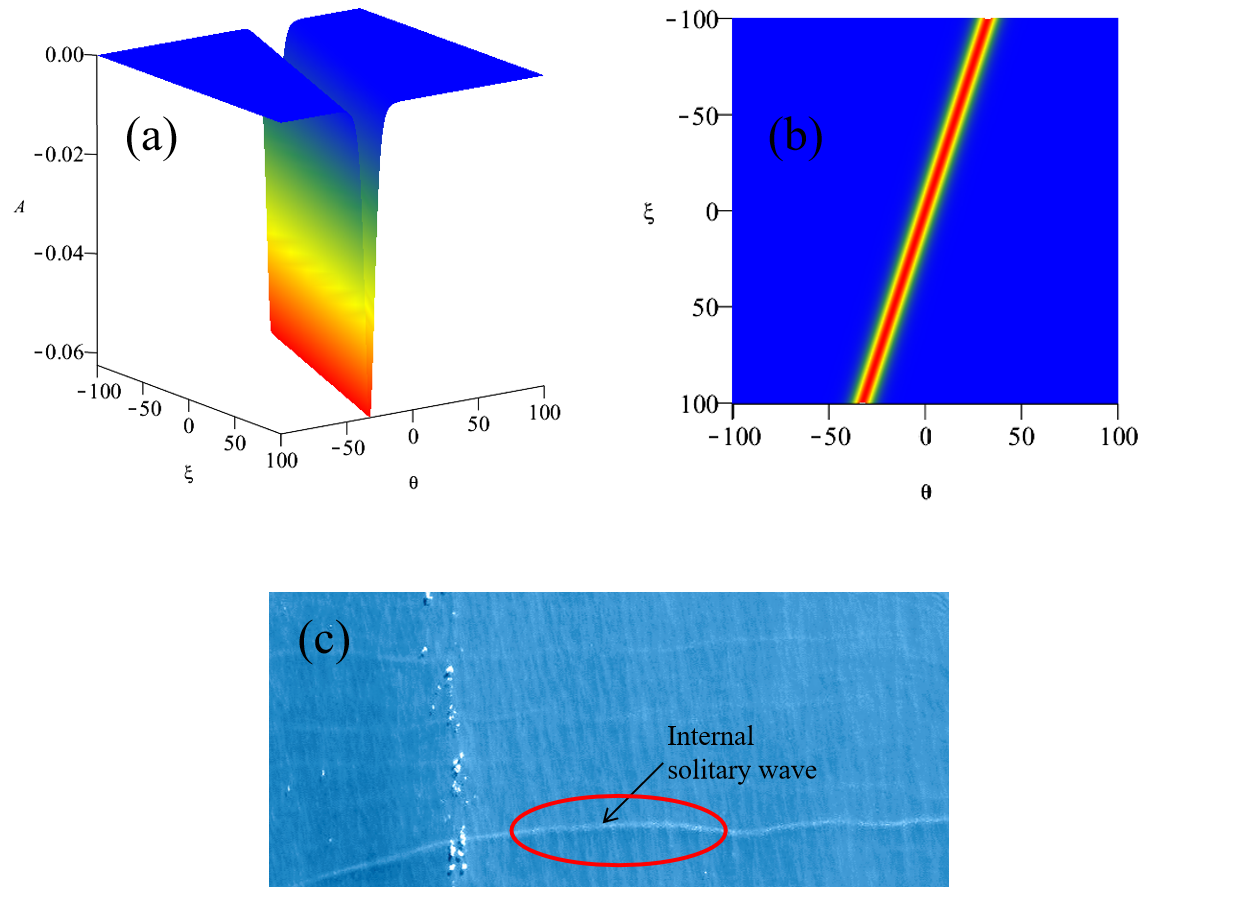}
     \caption{(a) The ``dark"  1-soliton solution of Eq. \eqref{76}, with $H=8$, $D=5$, $d=0.3$, $U_{1}=0$, $U_{2}=0.4$, $U_{3}=1.8$, $\frac{\rho_{1}}{\rho_{3}}=1.5$, $\left(k_1, k_2\right)=(-0.15,0.1)$ and $\omega_j^0=0$ for $j=1, 2$ at $\tau=0$. (b) The density plot of the ``dark"  1-soliton solution. (c) The oceanic ISW (ASTER false-color VNIR
image over the area between the Andaman Sea and the Strait of Malacca acquired on 31 January 2002 at 0406 UTC), from \cite{iw}.}
  \label{fig-9}
       \end{center}
\end{figure}

Similar to the nonlinear coefficient $a_{1}$, Fig. \ref{fig-6} shows the variation of  the densities and shear flows can affect the range of $d$, and this effect considerably narrows the area of  $d$, which makes the middle layer thicker.
It is also the density that has a greater effect on the range of $d$, than  the shear flows. The value of the middle layer must be within a reasonable range.

The value of the coefficient $a_{3}$ is a tiny number (see Fig. \ref{fig-7}) compared to $a_{1}$ and $a_{2}$, mainly because the depth of the middle layer $d$ is  small compared to $H$ and $D$, and the upper and lower layers do not contribute to the value of $a_{3}$. The coefficient $a_{3}$ is zero at $D=0$. In addition, there is a zero point in the positive direction of the $D$-axis. Moreover, the position of this zero point is consistent with that of the singularity in Fig. \ref{fig-3}. Therefore, the change of the density ratio does not affect the position of the zero point, while the existence of the shear flows does. Specifically, increasing the shear flows makes the position of the zero point  move along the positive direction of the $D$-axis. This is also consistent with our previous analysis that the dispersion effect becomes weak at the position where the nonlinear effect is suddenly enhanced.

The range of $d$ is affected by the densities and shear flows, as depicted in Fig. \ref{fig-8}. The details are similar to the analysis of the nonlinear and dispersion terms and will not be stated again. Moreover, we note that $a_{3}$  approaches zero in the $\rho_{1}/\rho_{3}=1$ case, which reduces the KP equation to a KdV model.

When choosing the stratification location, one should keep the middle layer thin and try to avoid those stratification locations that make the coefficients tend to infinity or  zero. It is discovered that fixing $H$ and $c$ to be different values will lead to similar conclusions.

\section{Internal solitary wave interactions}

By rescaling the function and its variables as

\begin{equation}\label{82}
\begin{aligned}
A(\xi, \theta, \tau)=\frac{6}{a_{1}}u(\xi, \theta, \tau), \quad \tau=-\frac{\sqrt{a_{2}}\hat{\tau}}{4}, \quad \xi=\sqrt{a_{2}}\hat{\xi}, \quad \theta=\sqrt{\frac{a_{2}a_{3}}{3}}\hat{\theta},
\end{aligned}
\end{equation}
and then dropping the hats for convenience, Eq. \eqref{76} becomes
\begin{equation}
\begin{aligned}\label{83}
\left(-4u_\tau+6 u u_\xi+u_{\xi \xi \xi}\right)_\xi+3 u_{\theta \theta}=0.
\end{aligned}
\end{equation}

Due to the physical constraints, the coefficients $a_{2}$ and $a_{3}$ cannot be negative, thus,  Eq. \eqref{76}  can only be transformed to the KP-II equation. This  indicates the absence of the (2+1)-dimensional internal rogue waves  described by Eq. \eqref{76}, and actually, the current research on internal rogue waves mainly relies on the (1+1)-dimensional Gardner equation\cite{16}. In the following, solutions of Eq. \eqref{76} are obtained from those of  Eq. \eqref{83},  and then  are used to investigate the internal solitary wave interactions.

\begin{figure}[!htb]
  \begin{center}
     \includegraphics[width=15cm]{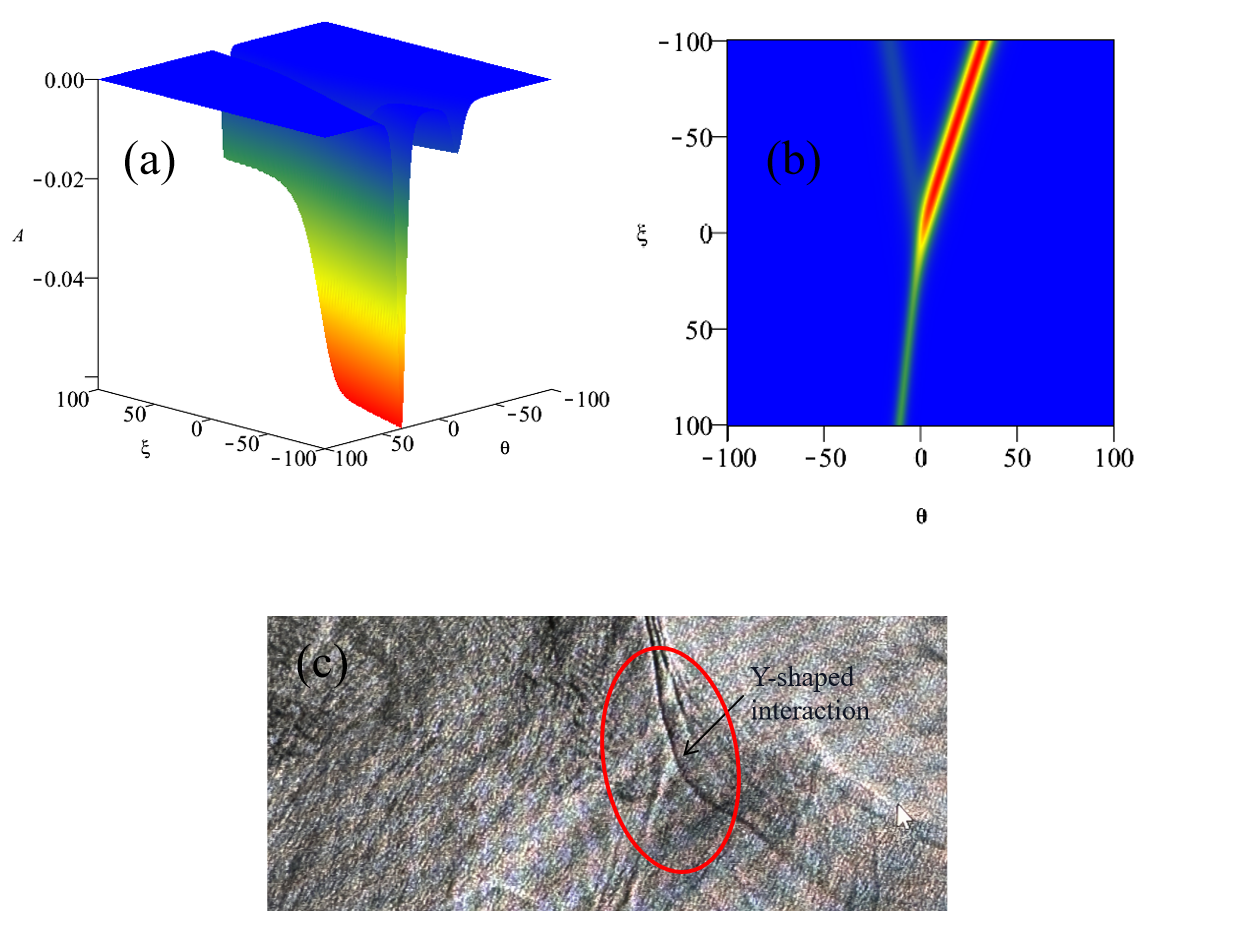}
     \caption{(a) The ``dark" Y-shaped solution of Eq. \eqref{76}, with $H=8$, $D=5$, $d=0.3$, $U_{1}=0$, $U_{2}=0.4$, $U_{3}=1.8$, $\frac{\rho_{1}}{\rho_{3}}=1.5$, $\left(k_1, k_2, k_3\right)=(-0.15,0,0.1)$  and $\omega_j^0=0$ for $j=1, 2, 3$ at $\tau=0$. (b) The density plot of the ``dark" Y-shaped solution. (c) The interaction of oceanic ISW (RADARSAT-1  image showing internal waves off the coast of Washington State, acquired 9 August 1999 at 0155 UTC), from \cite{iw}.}
  \label{fig-91}
       \end{center}
\end{figure}

\subsection{Review of  solutions of the KP equation (4.2)}
Solutions of Eq. \eqref{83} can be given as

\begin{equation}\label{84}
\begin{aligned}
u(\xi,\theta, \tau)=2 \frac{\partial^2}{\partial x^2} \ln \lambda (\xi,\theta, \tau),
\end{aligned}
\end{equation}
where $\lambda (\xi,\theta, \tau)$ can be expressed in terms of the Wronskian determinant
\begin{equation}\label{85}
\lambda=\operatorname{Wr}\left(f_1, \ldots, f_N\right)=\left|\begin{array}{ccc}
f_1^{(0)} & \cdots & f_N^{(0)} \\
\vdots & \ddots & \vdots \\
f_1^{(N-1)} & \cdots & f_N^{(N-1)}
\end{array}\right|,
\end{equation}
with $f_i^{(n)}=\partial^n f_i / \partial \xi^n $,  $f_i$ being the set of linearly independent solutions of  $\frac{\partial f_i}{\partial \theta}=\frac{\partial^2 f_i}{\partial \xi^2}$ and $\frac{\partial f_i}{\partial \tau}=\frac{\partial^3 f_i}{\partial \xi^3}$.

The $N$-soliton solution is obtained by taking
\begin{equation}\label{86}
f_i=\sum_{j=1}^M a_{i j} \mathrm{e}^{\omega_j}, \quad \text { for } i=1, \ldots, N, \quad \text { and } \quad M>N \text {, }
\end{equation}
where the constants $a_{i j}$  define the $N \times M$ coefficient matrix $C_{(N, M)}=\left(a_{i j}\right)$. The phase functions $\omega_j$  can be written in the  form of
\begin{equation}\label{87}
\omega_j(\xi, \theta, \tau)=-k_j \xi+k_j^2 \theta-k_j^3 \tau+\omega_j^0, \quad \text { for } \quad j=1, \ldots, M,
\end{equation}
where $k_j$ and $\omega_j^0$ are arbitrary constants, and note that $k_1<k_2<\cdots<k_M$.

\begin{figure}[!htb]
  \begin{center}
     \includegraphics[width=15cm]{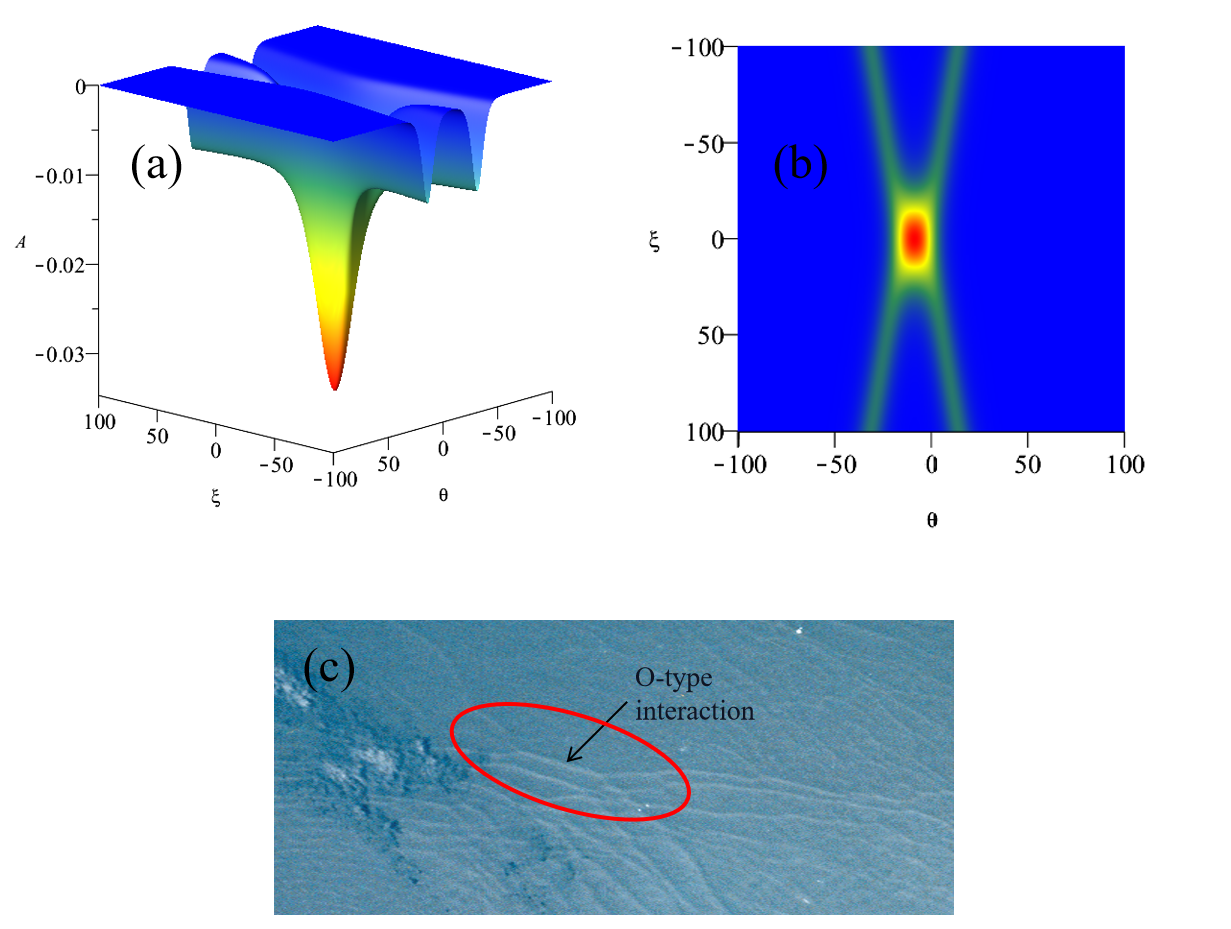}
     \caption{(a) The ``dark" ordinary 2-soliton solution (O-type) of Eq. \eqref{76} with $H=8$, $D=5$, $d=0.3$, $U_{1}=0$, $U_{2}=0.4$, $U_{3}=1.8$, $\frac{\rho_{1}}{\rho_{3}}=1.5$, $\left(k_1, \ldots, k_4\right)=(-0.1,-0.001,0,0.1)$ and $\omega_j^0=0$ for $j=1, \ldots, 4$ at $\tau=1$. (b) The density plot of the ``dark" O-type solution. (c) The interaction of oceanic ISW (Astronaut photograph (STS036-082-76) acquired on 1 March 1990 at 1254 UTC), from \cite{iw}.}
  \label{fig-10}
       \end{center}
\end{figure}

By choosing the appropriate forms of  $C_{(N, M)}$, some exact solutions of  Eq. \eqref{83} can be obtained. For the simplest example with $N=1$ and $M=2$, i.e. $\tau=f_1=a_{11} \mathrm{e}^{\omega_1}+a_{12} \mathrm{e}^{\omega_2}$ with $a_{11} a_{12}>0$, we obtain the  1-soliton solution

\begin{equation}
\begin{aligned}\label{88}
u=\frac{2 \mathrm{e}^{-k_1^3 \tau-k_2^3 \tau+k_1^2 \theta+k_2^2 \theta-k_1 \xi-k_2 \xi}(k_1-k_2)^2}{\left(\mathrm{e}^{-k_1^3 \tau+k_1^2 \theta-k_1 \xi}+\mathrm{e}^{-k_2^3 \tau+k_2^2 \theta-k_2 \xi}\right)^2}.
\end{aligned}
\end{equation}
Similarly, let $N=1$ and $M=3$, the Y-shaped solution  whith three line solitons interacting at a vertex is obtained.

It is well known that elastic 2-soliton solutions\cite{34} of Eq. \eqref{83}  have been   classified into three  types: ordinary (O-type), asymmetric (P-type) and resonant (T-type).  These types are generated by choosing $N=2$ and $M=4$, and their corresponding coefficient matrices have the following forms, respectively,
\begin{equation}\label{89}
C_{\mathrm{O}}=\left(\begin{array}{cccc}
1 & 1 & 0 & 0 \\
0 & 0 & 1 & 1
\end{array}\right), \quad C_{\mathrm{P}}=\left(\begin{array}{cccc}
1 & 0 & 0 & -1 \\
0 & 1 & 1 & 0
\end{array}\right), \quad C_{\mathrm{T}}=\left(\begin{array}{cccc}
1 & 0 & - & - \\
0 & 1 & + & +
\end{array}\right),
\end{equation}
where $'+,-'$ indicates the sign of the non-zero entry.

\subsection{The internal  solitary wave interactions}

According to Eq. \eqref{82},   the 1-soliton , Y-shaped , ordinary 2-soliton, asymmetric 2-soliton , and  resonant  2-soliton solutions  for  Eq. \eqref{83} can be used to build  solutions of Eq. \eqref{76}. For instance,  taking the simplest example, from  Eq. \eqref{88}, we obtain

\begin{equation}
\begin{aligned}\label{90}
A=\frac{12 \mathrm{e}^{\frac{4 k_1^3 \tau}{\sqrt{a_2}}+\frac{4 k_2^3 \tau}{\sqrt{a_2}}+k_1^2  \sqrt{\frac{3}{a_2 a_3}} \theta+k_2^2 \sqrt{\frac{3}{a_2 a_3}} \theta-\frac{k_1 \xi}{\sqrt{a_2}}-\frac{k_2 \xi}{\sqrt{a_2}}}(k_1-k_2)^2}{a_1\left(e^{\frac{4 k_1^3 \tau}{\sqrt{a_2}}+k_1^2  \sqrt{\frac{3}{a_2 a_3}} \theta-\frac{k_1 \xi}{\sqrt{a_2}}}+\mathrm{e}^{\frac{4 k_2^3 \tau}{\sqrt{a_2}}+k_2^2  \sqrt{\frac{3}{a_2 a_3} }} \theta-\frac{k_2 \xi}{\sqrt{a_2}}\right)^2}.
\end{aligned}
\end{equation}

In order to determine the specific values of the coefficients $a_{1}$, $a_{2}$ and $a_{3}$, we set $D=5$, $H=8$  and $d=0.3$ so that the lower layer is  deep  and  the middle  is  thin. Under different densities and shear flows, we can determine the values of the coefficients from Figs.  \ref{fig-3}-\ref{fig-8}.

An oceanic internal solitary wave can be well described by the ``dark"  1-soliton solution, as depicted in Fig. \ref{fig-9}. The ``dark" Y-shaped solution (see Fig. \ref{fig-91})  is formed by the resonant interaction of three oceanic ISW at a vertex, which demonstrates that the interaction of ISW can produces a Miles resonance. Miles resonance can be regarded as one of the basic structures of the resonance interaction of elastic two solitons.
\begin{remark}
The Miles resonance  corresponds to an internal solitary wave pattern  captured along the coast of Washington State in $1990$ by the RADARSAT-1 satellite, as shown in Fig. \ref{fig-91} (c). Unlike the eastern coast of the United States, the western coast lacks an extensive continental shelf, leading to the occurrence of these internal solitary waves closer to the shore.
\end{remark}

\begin{figure}[!htb]
  \begin{center}
     \includegraphics[width=15cm]{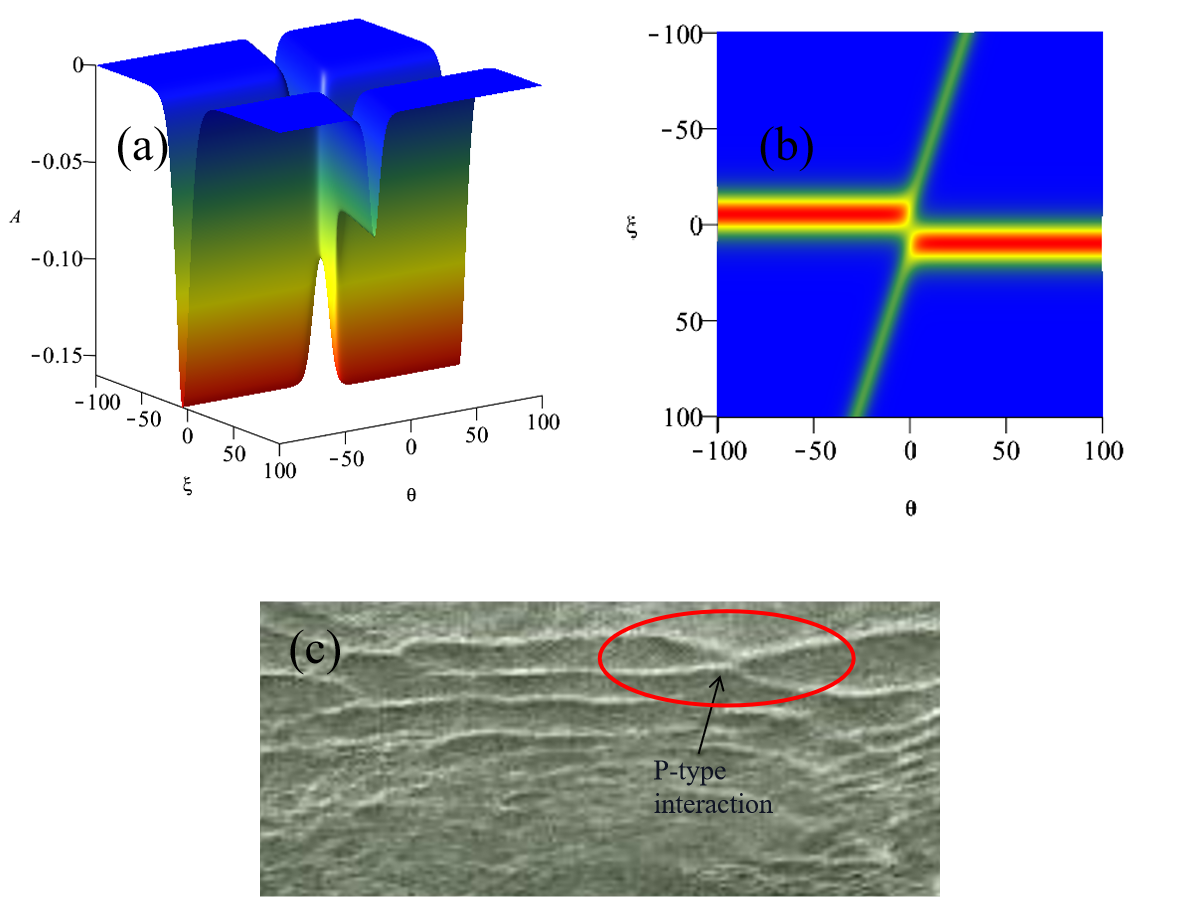}
     \caption{(a) The ``dark" asymmetric 2-soliton solution (P-type) of Eq. \eqref{76} with $H=8$, $D=5$, $d=0.3$, $U_{1}=0$, $U_{2}=0.4$, $U_{3}=1.8$, $\frac{\rho_{1}}{\rho_{3}}=1.5$, $\left(k_1, \ldots, k_4\right)=(-0.2,-0.15,0.1,0.2)$ and $\omega_j^0=0$ for $j=1, \ldots, 4$ at $\tau=0$. (b) The density plot of the ``dark" P-type solution. (c) The interaction of oceanic ISW (ERS-2 SAR
image of the Andaman acquired on 11 February 1997 at 0359 UTC), from \cite{iw}.}
  \label{fig-11}
       \end{center}
\end{figure}

\begin{figure}[!htb]
  \begin{center}
     \includegraphics[width=15cm]{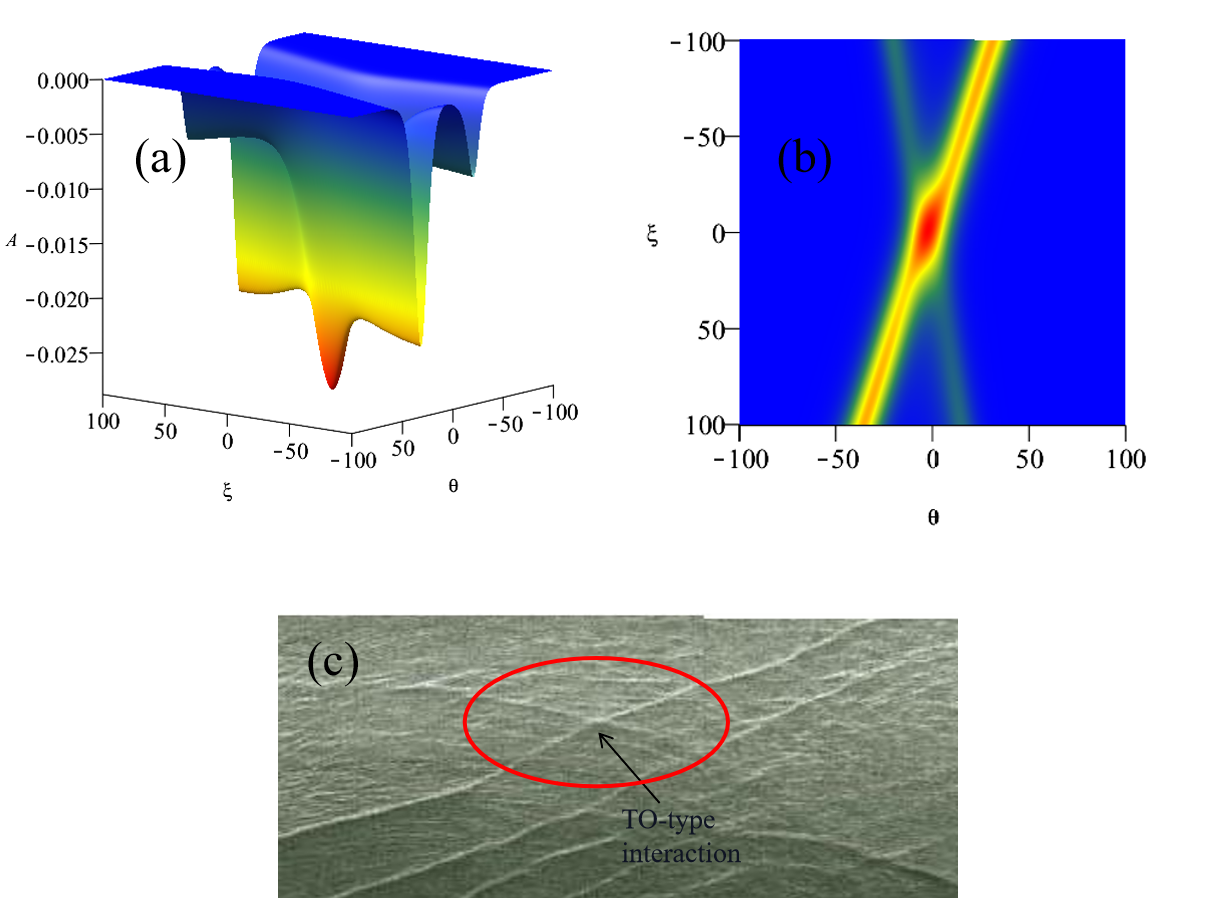}
     \caption{(a) The ``dark" asymmetric 2-soliton solution (TO-type) of Eq. \eqref{76}, with $H=8$, $D=5$, $d=0.3$, $U_{1}=0$, $U_{2}=0.4$, $U_{3}=1.8$, $\frac{\rho_{1}}{\rho_{3}}=1.5$, $\left(k_1, \ldots, k_4\right)=(-0.1,0,0.05,0.1)$ and $\omega_j^0=0$ for $j=1, \ldots, 4$ at $\tau=3$. (b) The density plot of the ``dark" TO-type solution.
     (c) The interaction of oceanic ISW (ERS-2 SAR
image of the Andaman acquired on 11 February 1997 at 0359 UTC), from \cite{iw}.}
   \label{fig-12}
       \end{center}
\end{figure}

\begin{figure}[!htb]
  \begin{center}
     \includegraphics[width=15cm]{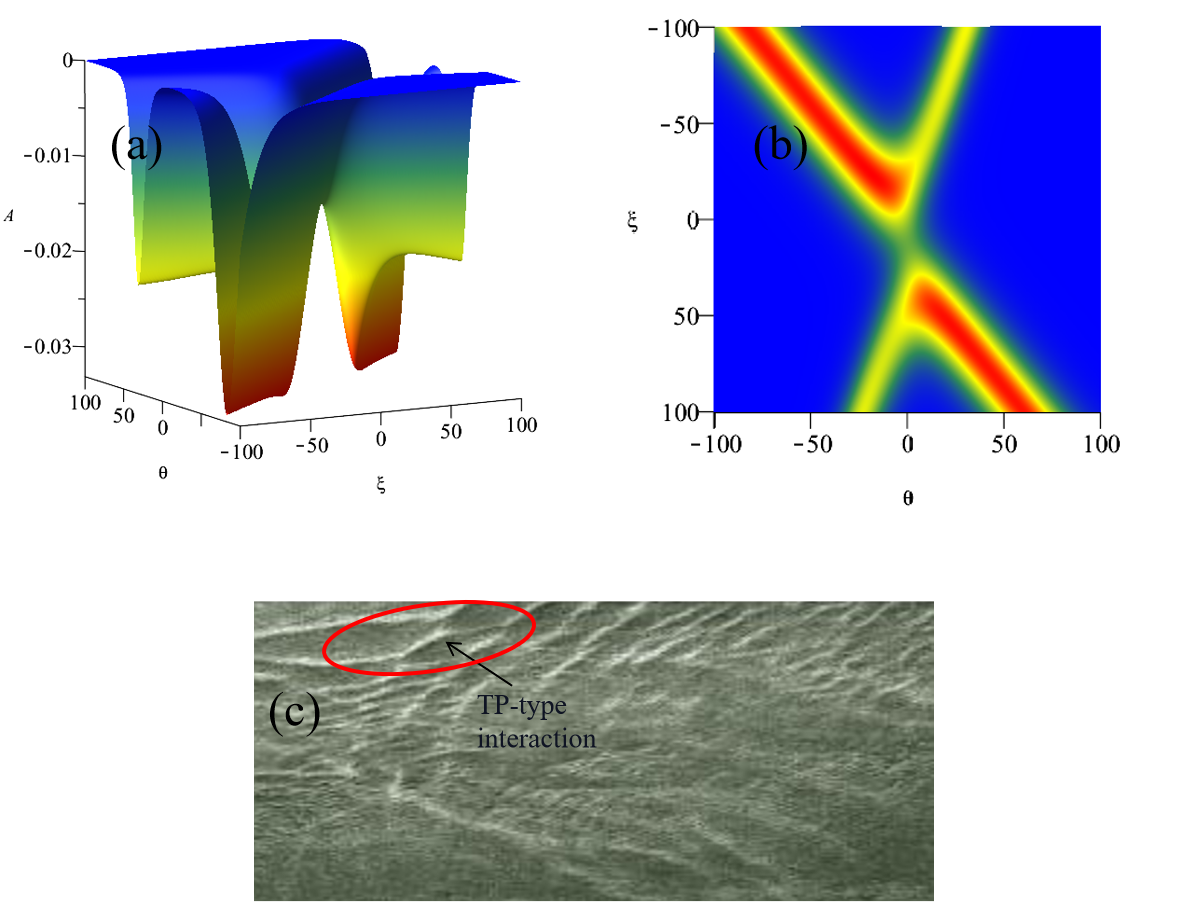}
     \caption{(a) The ``dark" asymmetric 2-soliton solution (TP-type) of Eq. \eqref{76}, with $H=8$, $D=5$, $d=0.3$, $U_{1}=0$, $U_{2}=0.4$, $U_{3}=1.8$, $\frac{\rho_{1}}{\rho_{3}}=1.5$, $\left(k_1, \ldots, k_4\right)=(-0.1,-0.08,0.05,0.1)$ and $\omega_j^0=0$ for $j=1, \ldots, 4$ at $\tau=3$.
     (b) The density plot of the ``dark" TP-type solution. (c) The interaction of oceanic ISW (ERS-2 SAR
image of the Andaman acquired on 11 February 1997 at 0359 UTC), from \cite{iw}.}
   \label{fig-121}
       \end{center}
\end{figure}

Now, we focus on the types of interactions of two oceanic ISW. Firstly, the ``dark"  O-type solution of Eq. \eqref{76} is obtained from the ordinary 2-soliton solution of  Eq. \eqref{83} through Eq. \eqref{82}. As can be seen from Fig. \ref{fig-10}, the ordinary  interactions of  the ISW produce a region where a wave with a relatively large amplitude exists. In this specific case, the amplitude of the wave in this region is more than twice that of a single internal solitary wave. The ISW produce a phase shift in this region. In real physical situations, the phase shift is not very large, usually twice the wavelength of the soliton at most\cite{33}.
It is important to note that although we only show  the figures of the O-type interactions at a certain moment, in fact, the size of the region neither expands nor contracts with time, and the amplitude of the wave in the region is also stable. It follows that the  interactions of  the ISW produce a wave with a relatively large amplitude, which propagates  without taking into account the frictional dissipation.
\begin{remark}
The ordinary interactions (O-type) align with Fig. \ref{fig-10} (c) were captured by the RADARSAT-1 satellite in the South African maritime region in $1990$.
Zheng et al. analyzed the image and found that both sets of waves propagate toward the shore, complex wave-wave interactions occur when the two sets of waves meet, and that the water depths of ISW at this site are all less than $500$ $m$, with intervals ranging from $1.08\thicksim2.27$ $km$, and peak lengths ranging from $50\thicksim100$ $km$ \cite{z}.
\end{remark}

Secondly, we display the P-type interactions for the internal waves in Fig. \ref{fig-11}. The difference from the ordinary interactions is that the amplitudes of the two solitary waves are different, and the amplitude of the asymmetric interaction region is always smaller than that of the highest soliton.
Also we note that the solitons with the largest amplitude are almost parallel to the $\theta$-direction.

\begin{figure}[!htb]
  \begin{center}
     \includegraphics[width=15cm]{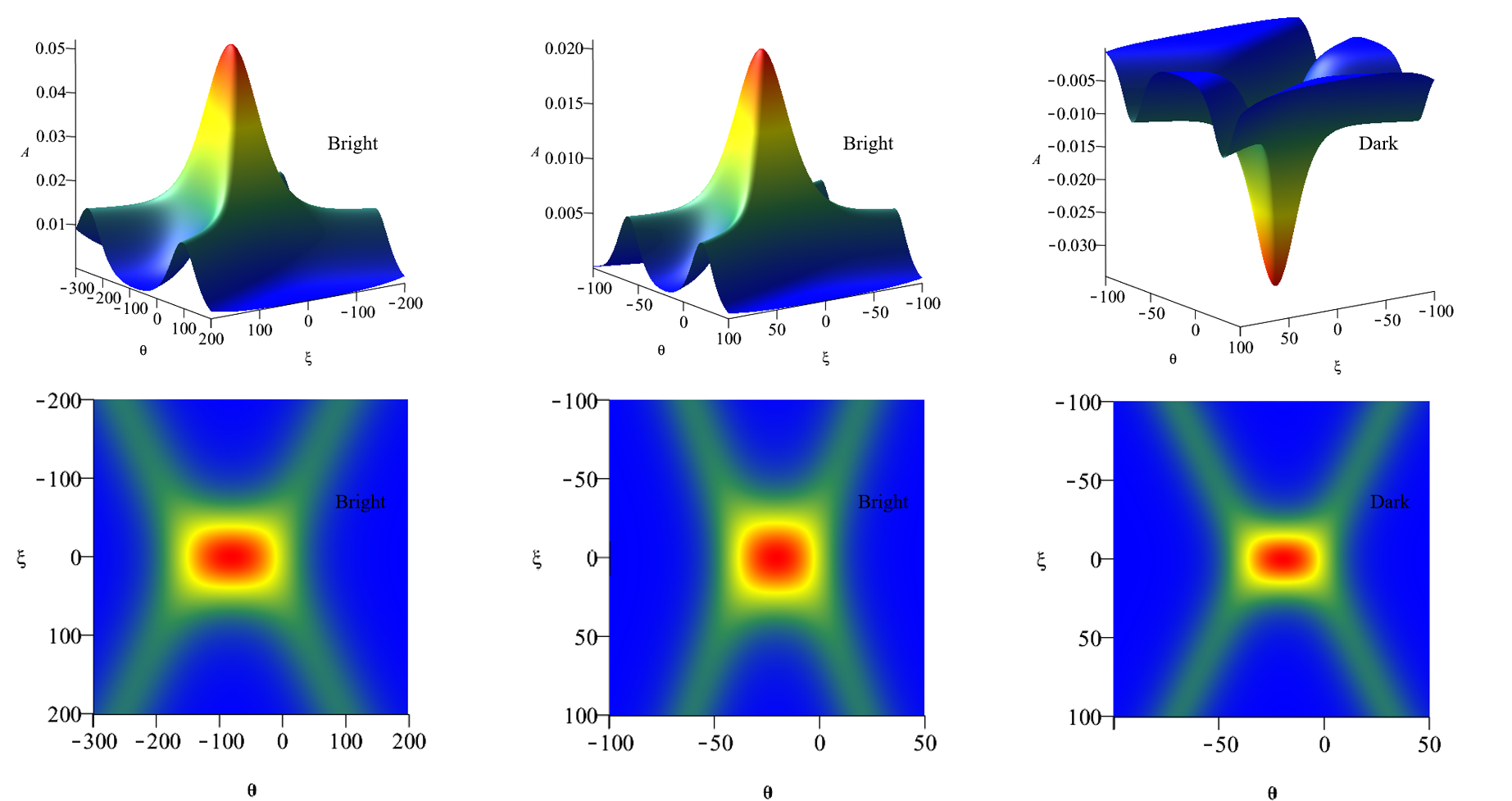}
     \caption{Left: The ``bright" ordinary 2-soliton solution  of Eq. \eqref{76} with $U_{2}=0$ and $U_{3}=0$. Center: The ``bright" ordinary 2-soliton solution  of Eq. \eqref{76} with $U_{2}=0.2$ and $U_{3}=0.9$. Right: The ``dark" ordinary 2-soliton solution of Eq. \eqref{76} with  $U_{2}=0.4$ and $U_{3}=1.8$. In all cases $\left(k_1, \ldots, k_4\right)=(-0.1,-0.001,0,0.1)$, $U_{1}=0$, $H=8$, $D=5$, $d=0.3$, $\frac{\rho_{1}}{\rho_{3}}=2$ and $\omega_j^0=0$ for $j=1, \ldots, 4$ at $\tau=1$.}
  \label{fig-13} %% label for first subfigure
       \end{center}
\end{figure}

Thirdly, making advantages of  the resonant 2-soliton solution of Eq. \eqref{83} and the scaling of the variables \eqref{82}, we can obtain the solution of Eq. \eqref{76}  to discuss whether there are resonant interactions, i.e., T-type interactions (web-solition) in the interior of the fluid.
As displayed in Fig. \ref{fig-12}, though we do not find resonant interactions of  two ISW,  we obtain another asymmetric interaction (TO-type).
While this interaction shares some similarities with the O-type interaction, the TO-type interaction is distinct in that it is generated by two internal waves  with varying amplitudes. It is evident that the amplitude of the interaction region does not exceed several times that of the higher soliton's amplitude, and the phase shift is not significant

%Actually, the interaction pattern marked at the bottom of Fig. 2 provides the evidence for the actual existence of the TO-type interactions.

\begin{figure}[!htb]
  \begin{center}
     \includegraphics[width=15cm]{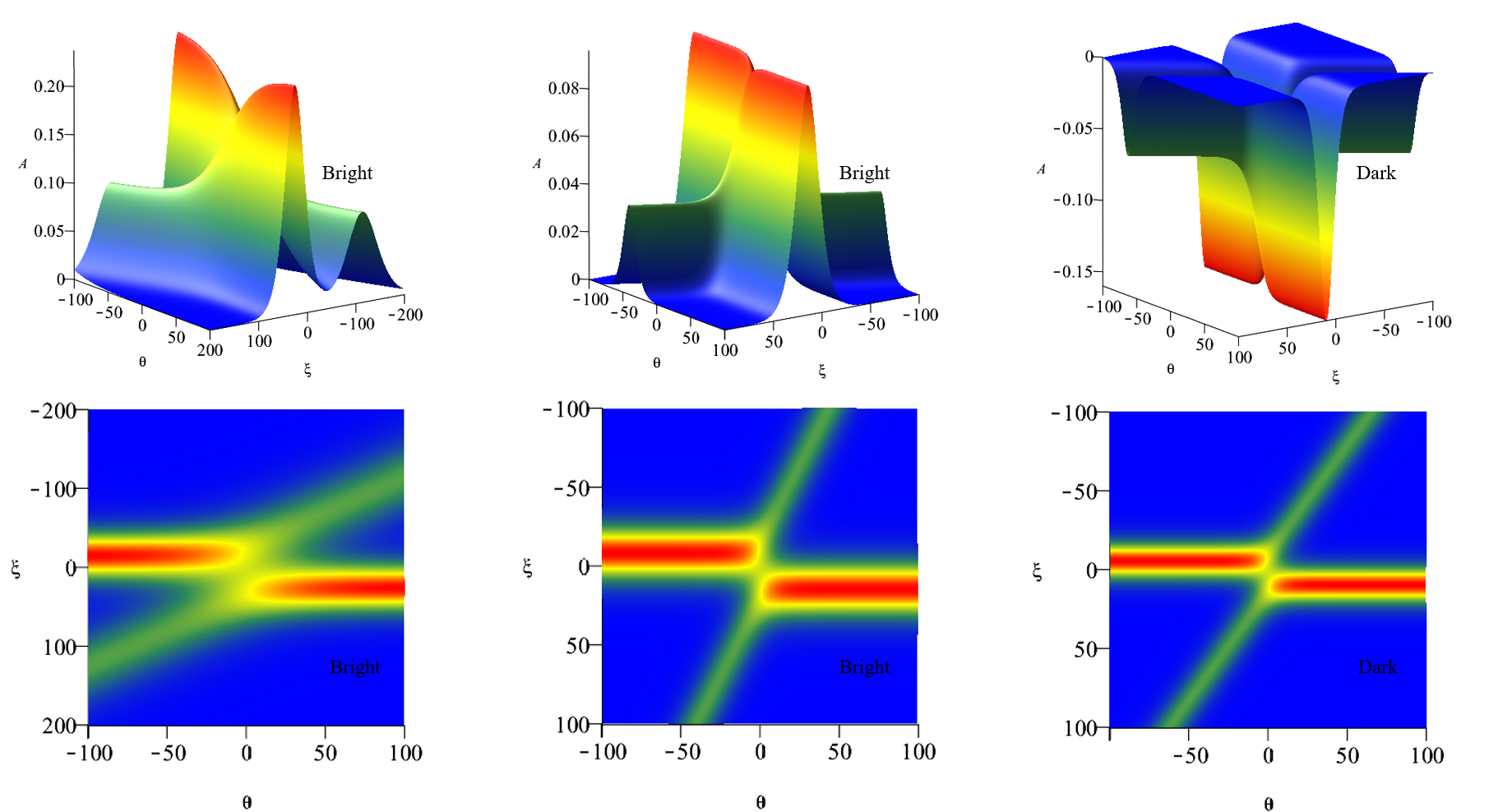}
     \caption{Left: The ``bright" asymmetric 2-soliton solution  of Eq. \eqref{76} with  $U_{2}=0$ and $U_{3}=0$. Center: The ``bright" asymmetric 2-soliton solution  of Eq. \eqref{76} with  $U_{2}=0.2$ and $U_{3}=0.9$. Right: The ``dark" asymmetric 2-soliton solution  of Eq. \eqref{76} with  $U_{2}=0.4$ and $U_{3}=1.8$. In all cases $\left(k_1, \ldots, k_4\right)=(-0.2,-0.15,0.1,0.2)$,  $U_{1}=0$, $H=8$, $D=5$, $d=0.3$, $\frac{\rho_{1}}{\rho_{3}}=2$ and $\omega_j^0=0$  for $j=1, \ldots, 4$ at $\tau=0$.}
  \label{fig-14} %% label for first subfigure
       \end{center}
\end{figure}

When varying the values of $k_{1}, k_{2}, k_{3}$ and $k_{4}$ (note that $k_1<k_2<k_3<k_4$),
a third asymmetric interaction, referred to as the TP-type interaction, is revealed. It shares some similarities with the P-type interaction, but in contrast to the P-type interaction, the amplitude of this TP-type interaction region becomes lower than that of any individual internal solitary wave. This  characteristic results in a less conspicuous interaction region when observed in satellite imagery. Moreover, the soliton with a higher amplitude is notably no longer aligned parallel to the $\theta$-direction.
\begin{remark}
The asymmetric interactions (P-type, TO-type, and TP-type) exhibit features consistent with Figs. \ref{fig-11} (c)- \ref{fig-121} (c) observed in satellite imagery from the Andaman Sea in $1997$, acquired by the ERS-2 satellite equipped with SAR.
Alpers et al.  identified several sources of internal waves based on images of the region (The shallow ridges between the Nicobar and
Andaman islands, submarine banks,  and the shallow reefs off the northwest coast of Sumatra) \cite{an}.
\end{remark}

\begin{figure}[!htb]
  \begin{center}
     \includegraphics[width=15cm]{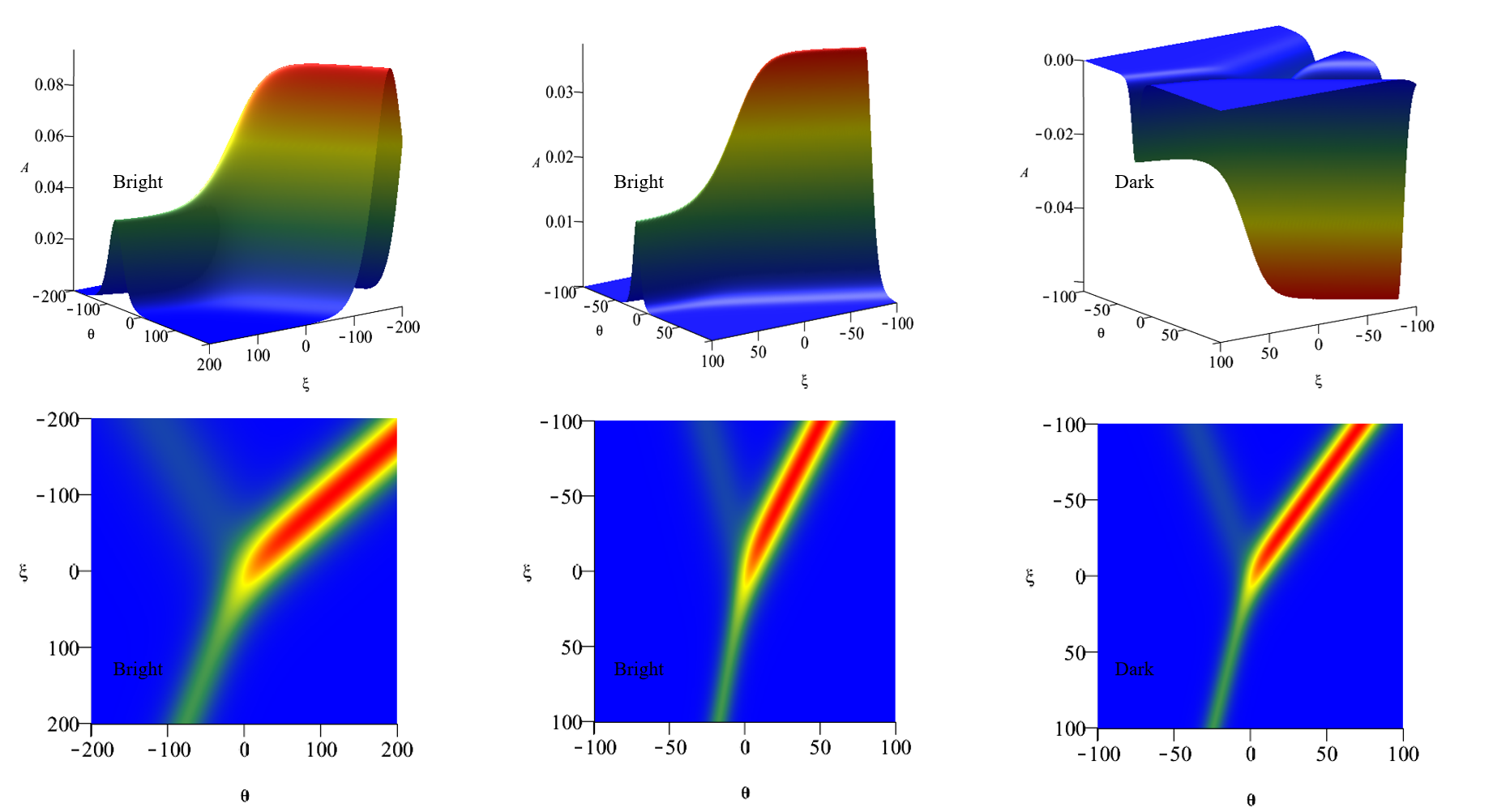}
     \caption{Left: The ``bright" Y-shaped solution  of Eq. \eqref{76} with  $U_{2}=0$ and $U_{3}=0$. Center: The ``bright" Y-shaped solution  of Eq. \eqref{76} with  $U_{2}=0.2$ and $U_{3}=0.9$. Right: The ``dark" Y-shaped solution  of Eq. \eqref{76} with  $U_{2}=0.4$ and $U_{3}=1.8$. In all cases $\left(k_1, \ldots, k_4\right)=(-0.2,-0.15,0.1,0.2)$,  $U_{1}=0$, $H=8$, $D=5$, $d=0.3$, $\frac{\rho_{1}}{\rho_{3}}=2$ and $\omega_j^0=0$  for $j=1, \ldots, 4$ at $\tau=0$.}
  \label{fig-151} %% label for first subfigure
       \end{center}
\end{figure}

The above results demonstrate that the resonance 2-soliton solution behaves as asymmetric interactions (TO-type or TP-type) for the oceanic internal waves. That is, ISW  exhibit two major types of fundamental interactions, ordinary interactions and asymmetric interactions including TO-type and TP-type, and the resonance interactions show Miles resonance of three ISW.

Finally, we study the influence of the densities and shear flows on the  interactions of the oceanic ISW.  As shown in Fig. \ref{fig-13}, in the  absence of shear flows or the presence of  relatively small shear flows, we obtain the ``bright" ordinary 2-soliton solutions, while when there are relatively large shear flows, the ``dark" ordinary 2-soliton solution is produced.
This ``dark" ordinary 2-soliton solution corresponds to the generation of internal wave interactions, underscoring the crucial role of shear flows in the formation of internal waves.
In fact, whether the ``bright" or the ``dark" soliton solution is obtained depends on the sign of the nonlinear coefficient $a_{1}$, see Fig. \ref{fig-3} where the shear flows  affect the sign of  $a_{1}$ once  the stratification  and the ratio of density are determined. In addition, the shear flows affect the amplitude and size of the ordinary interaction region, both  do not vary with time. Comparing Figs. \ref{fig-10} and \ref{fig-13}, it can be seen that increasing the ratio of the densities  has little effect on the amplitude of the interaction region, but changes the size of the region.

In the case of asymmetric interactions (see Fig. \ref{fig-14}), no shear flows or  relatively small shear flows generate the ``bright" asymmetric 2-soliton solutions, and relatively large shear flows excite a ``dark" asymmetric 2-soliton solution (the emergence of internal wave interactions).  Similar to the case in Fig. \ref{fig-11}, this is all due to the action of the shear flows.
It can be observed from Fig. \ref{fig-14} with Fig. \ref{fig-11} that increasing the ratio of the densities  has almost no effect on the amplitude of the asymmetric interaction region, but changes the size of the region.
The case of Miles resonance(see Figure \ref{fig-151}) is similar and will not be repeated here.

The effect of shear flows on the ordinary, asymmetric and Miles resonance interactions of the internal waves is similar.
Here we take only three types of typical interaction 3D images as examples. As a side note, the analysis of the TO-type and TP-type interactions also leads  to the same conclusion.
In Figs. \ref{fig-13}-\ref{fig-151}, we have coarsely analyzed the effects of shear flows and density on the internal solitary wave interactions. In order to find out the rules, we next study their effects more pertinently.

\begin{table}[htbp]
\caption{Comparison of the maximum amplitude of the  solitary wave-wave interactions for the derived KP equation and the KP-II equation.}
\label{table3-3}
\centering
\begin{tabular}{c|c|c|c|c|c|c}
\bottomrule
\diagbox{\textbf{\textbf{Equation}}}{\textbf{Type}}  & 1-soliton & O-type & P-type & Y-shape & TO-type & TP-type  \\ \hline
KP      & 0.094         & 0.052    & 0.24     & 0.094         & 0.043    & 0.05         \\ \hline
KP-II  & 0.031 & 0.016      &  0.079     & 0.031         & 0.014(T)    & 0.017(T)    \\ \toprule
\end{tabular}
\end{table}

In Table 1, we compared the maximum amplitudes of the derived KP equation \eqref{76} and the KP-II equation \eqref{83}. In order to control the variables, it is ensured that the values of $k_j$ are the same and that there is no shear flow in Eq. \eqref{76}. Since Eq. \eqref{83} usually yields interactions above the zero background, we use the ``bright" interactions of Eq. \eqref{76} as comparison. The T-type interaction of  Eq. \eqref{83} is used as  comparison between the TO-type and TP-type interactions of Eq. \eqref{76}. The amplitudes of all types of interactions obtained in Eq. \eqref{76} are much higher (about three times higher) than the corresponding interactions in Eq. \eqref{83}. This indicates that Eq. \eqref{76} yields  solitary wave-wave interactions with larger amplitudes.

\begin{table}[htbp]
\caption{Comparison of the maximum amplitude of the internal  solitary wave interactions under different shear flows, with the ``-" sign indicating the appearance of ``dark" interactions.}
\label{table3-3}
\centering
\begin{tabular}{c|c|c|c|c|c|c}
\bottomrule
\diagbox{\textbf{\textbf{Shear flow}}}{\textbf{Type}}  & 1-soliton & O-type & P-type & Y-shape & TO-type & TP-type  \\ \hline
$U_{1}=0$, $U_{2}=0$, $U_{3}=0$      & 0.094         & 0.052    & 0.24     & 0.094         & 0.043    & 0.05         \\ \hline
$U_{1}=0$, $U_{2}=0.2$, $U_{3}=0.9$      & 0.037         & 0.021    & 0.096     & 0.037         & 0.017    & 0.02         \\ \hline
$U_{1}=0$, $U_{2}=0.4$, $U_{3}=1.8$ & -0.064 & -0.035      &  -0.16    & -0.062         & -0.029    & -0.033    \\ \toprule
\end{tabular}
\end{table}

\begin{figure}[!htb]
  \begin{center}
     \includegraphics[width=15cm]{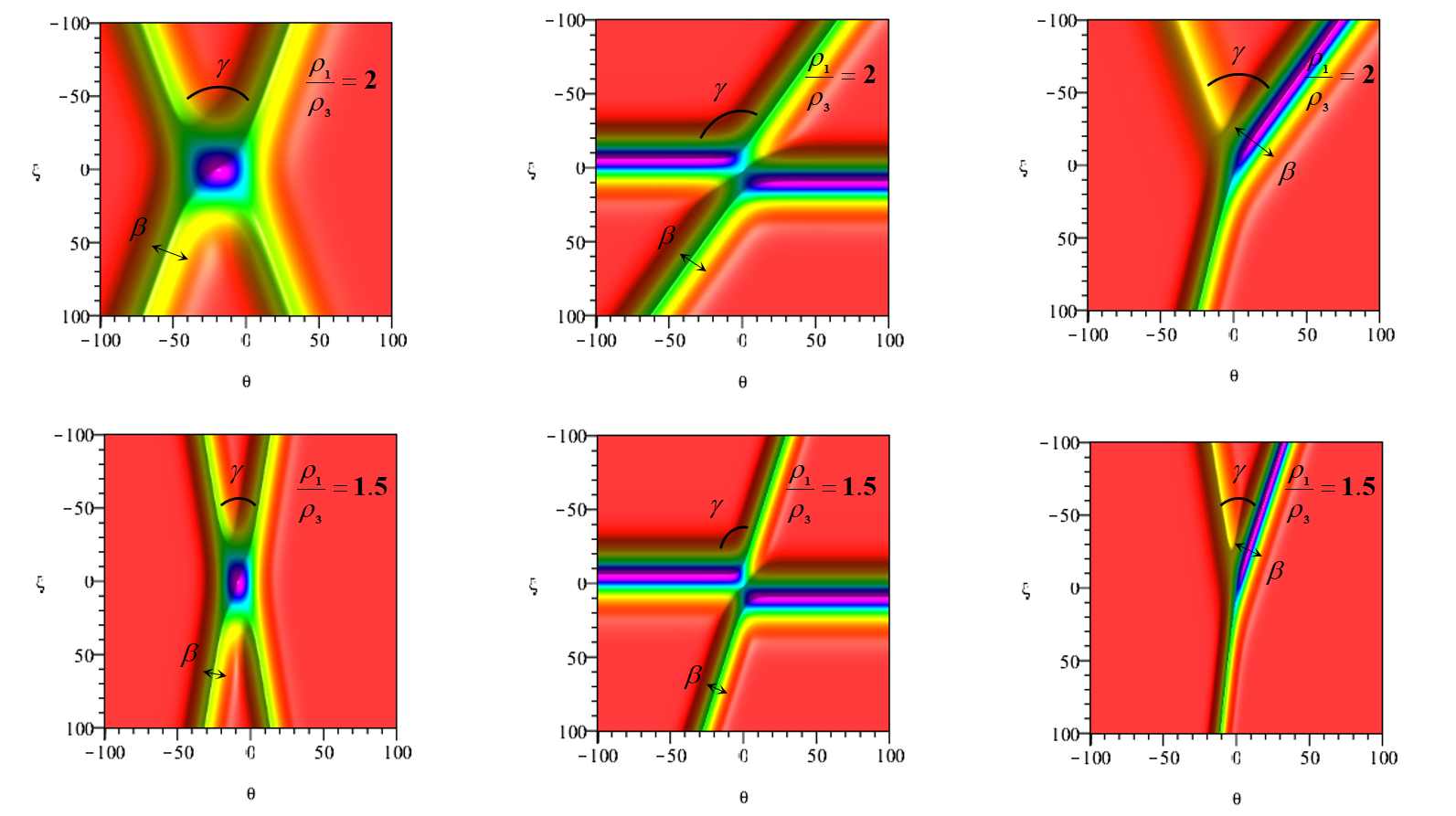}
     \caption{Left: The O-type solution  of Eq. \eqref{76} with $\left(k_1, \ldots, k_4\right)=(-0.1,-0.001,0,0.1)$ at $\tau=1$. Center: The P-type solution  of Eq. \eqref{76} with $\left(k_1, \ldots, k_4\right)=(-0.2,-0.15,0.1,0.2)$ at $\tau=0$. Right: The Y-shaped solution  of Eq. \eqref{76} with $\left(k_1, k_2, k_3\right)=(-0.15,0,0.1)$  at $\tau=0$. In all cases, $H=8$, $D=5$, $d=0.3$, $U_{1}=0$, $U_{2}=0.4$, $U_{3}=1.8$ and $\omega_j^0=0$ for $j=1, \ldots, 4$.}
  \label{fig-15}
       \end{center}
\end{figure}
Table 2 shows that with  the same density and stratification,  when the shear flow increases, the maximum amplitude of all types of  solitary wave-wave interactions decreases, and the ``bright" interactions will turn to ``dark" interactions.
Therefore, the presence of shear flows is the main determinant in exciting  ``bright" or ``dark"  solitary wave-wave interactions.
%in other words, shear flows determine the generation of internally isolated wave-wave interactions.
It is noted that ``dark" and ``bright" phenomena were also discussed in  \cite{16} for one-dimensional internal rogue waves governed by the Gardner equation, whereas only ``bright'' internal solitary wave interactions in the ocean were studied in \cite{30} due to the theory based on the constant coefficient.

The above results are obtained when a more realistic stratification is chosen $D=5$, i.e. the lower layer is a deep layer. When other reasonable stratifications are considered, one can judge the interactions are ``bright" or ``dark" from Figs. \ref{fig-3} and \ref{fig-4}. For example, when  the stratification is closer to the bottom of the fluid ($D=1$), the  interactions are ``dark"  without shear flows, and become ``bright" when increasing shear flows. When the stratification is near the middle of the fluid ($D=4$),  ``bright" interactions appear without shear flows, and as  shear flows increase ``dark" interactions come in being.

\begin{figure}[!htb]
  \begin{center}
     \includegraphics[width=12cm]{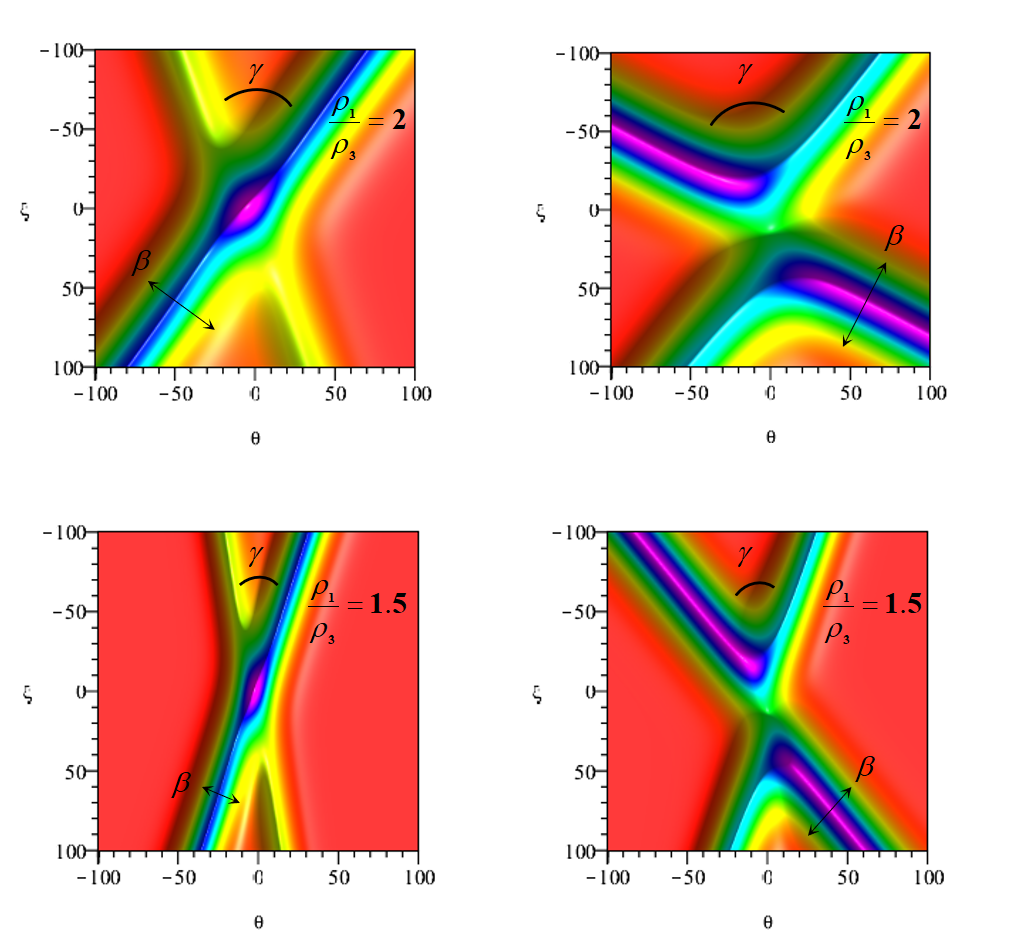}
     \caption{Left: The TO-type solution  of Eq. \eqref{76} with $\left(k_1, \ldots, k_4\right)=(-0.1,0,0.05,0.1)$. Right: The TP-type solution  of Eq. \eqref{76} with $\left(k_1, \ldots, k_4\right)=(-0.1,-0.08,0.05,0.1)$. In all cases, $H=8$, $D=5$, $d=0.3$, $U_{1}=0$, $U_{2}=0.4$, $U_{3}=1.8$, $\frac{\rho_{1}}{\rho_{3}}=1.5$ and $\omega_j^0=0$ for $j=1, \ldots, 4$ at $\tau=3$.
     }
  \label{fig-16}
       \end{center}
\end{figure}

The ratio of densities has no significant effect on the amplitudes of the internal solitary wave interactions, but has a fundamental impact on the angle between the ISW, the width of the waves, and the region of the interactions.
As shown in Fig. \ref{fig-15}, when the density ratio decreases from $\rho_{1}/\rho_{3}=2$ to $\rho_{1}/\rho_{3}=1.5$,
the angle $\gamma$ of  two ISW becomes smaller and the width $\beta$  also becomes narrower. In particular, the same phenomenon occurs in two special types of interactions as displayed in Fig. \ref{fig-16}. At the same time, as the ratio of densities decreases, the size of the area of interactions changes. In general, the change in angle $\gamma$ and width $\beta$ is the indirect cause of the size of the interacting area, but the fundamental factor is the change in the ratio of densities.

\begin{remark}
It is remarkable that one can transform the results into the laboratory coordinate system $(x, y, t)$, however, it will simply changes the scales of the spatial and temporal coordinates, and will not alter the main characteristics and properties of the internal solitary wave interactions that could be captured by the satellite images.
\end{remark}

%\newpage
\section{Conclusions and discussions}

We have established a (2+1)-dimensional KP model whose coefficients are functions of shear flow and density, considering a three-layer fluid with a continuous density distribution,  to investigate the oceanic  internal solitary wave interactions. These interactions are of various types in the ocean and take ``bright" and ``dark" forms under the influence of shear flow.
The analysis based on the coefficient shows that  the depth of the middle layer $d$ is smaller than that of the others. Besides, shear flow and density ratio  can affect the range of  $d$. In our specific case, we need to choose the value of $d$ carefully to ensure a thin intermediate layer. Simultaneously, when the delamination is located at a position where the nonlinear effect is suddenly enhanced, the dispersion effect is weakened and vice versa.

The internal solitary wave-wave interactions  are characterized by ordinary  and asymmetric interactions (which can  be  further classified  into four categories: O-type, P-type, TO-type and TP-type). The resonant interactions are manifested as the Miles resonance of three ISW. It is also  noted that  the resonance 2-soliton solution (web-soliton) can  evolve into TO-type or TP-type  interactions. Compared to common internal wave interactions, TO-type interactions have no significant displacement and TP-type interactions produce smaller amplitudes.
Different types of interactions show clear correspondences with internal wave satellite images. For instance, O-type interactions align with images from the southern African sea while asymmetric interactions like P-type, TO-type, and TP-type correspond with Andaman Sea satellite data. Moreover, Y-shaped interactions match those captured along the Washington State coast.

It is important to emphasize that we have simultaneously obtained both ``bright" and ``dark" forms of oceanic  internal solitary interactions. In fact, shear flow determines the emergence of ``bright" or ``dark" interactions in the ocean.
The density ratio has a significant effect on the angle, width and interaction area of ISW. The specific patterns of their influence are presented.
Furthermore, exploring new types of internal solitary wave interactions and different categories of internal waves, such as internal rogue waves and internal breathers, combined with the powerful tool of satellite imagery will be our primary focus in the future.

\appendix
\section{}
According to Eqs. \eqref{77}-\eqref{81},   exact expressions of the coefficients can be obtained as below.

\begin{flalign}
\begin{split}\label{97}
&I=\frac{\rho_1\left(-U_1+c\right)}{D}+\frac{\rho_3\left(-U_3+c\right)(H-D-d)}{\left(\sigma\left(-U_3+c\right)^2-H+D+d\right)^2}\\
&+\frac{1}{\sigma p_{1} d (-U_{2}+c)\left(e^{\frac{p_{1}}{\sigma(U_{2}-c)}}-\frac{1}{2}e^{\frac{2p_{1}}{\sigma(U_{2}-c)}}-\frac{1}{2}\right)}\cdot\\
&\left\{\operatorname { l n } \left( \frac { \rho _ { 1 } } { \rho _ { 3 } } \right) \left[\left(-p_1 d+p_2 \sigma\left(-U_2+c\right)\right) \rho_3 \sqrt{\frac{\rho_1}{\rho_3}} e^{\frac{p_1}{2 \sigma\left(U_2-c\right)}}\right.\right.\\
&-\left(p_1 d+p_2 \sigma\left(-U_2+c\right)\right) \rho_3 \sqrt{\frac{\rho_1}{\rho_3}} e^{\frac{3 p_1}{2 \sigma\left(U_2-c\right)}}\\
& +\frac{1}{4}\left(-U_2+c\right)\left(\left(\rho_1-\rho_3\right) p_1\left(-U_2+c\right)+p_2\left(\rho_1+\rho_3\right)\right) \sigma e^{\frac{2 p_1}{\sigma\left(U_2-c\right)}}\\
& -\frac{1}{2}p_1\left(\left(-U_2+c\right)^2\left(\rho_1-\rho_3\right) \sigma-2 d\left(\rho_1+\rho_3\right)\right) \sigma e^{\frac{2 p_1}{\sigma\left(U_2-c\right)}}\\
& \left.\left.-\frac{1}{4}\left(-\left(\rho_1-\rho_3\right)\left(-U_2+c\right) p_1+p_2\left(\rho_1+\rho_3\right)\right)\left(-U_2+c\right) \sigma \right]\right\}, \\
&
\end{split}&
\end{flalign}

\begin{flalign}
\begin{split}\label{98}
& \frac{2}{3} I \cdot a_1=\frac{\rho_1\left(-U_1+c\right)^2}{D^2}+\frac{\rho_3\left(-U_3+c\right)^2(H-D-d)}{\left(\sigma\left(-U_3+c\right)^2-H+D+d\right)^2} \\
& -\frac{1}{4 \sigma p_4 d^2 \rho_3 \sqrt{\frac{\rho_1}{\rho_3}}\left(e^{\frac{p_1}{\sigma\left(U_2-c\right)}}-e^{\frac{2 p_1}{\sigma\left(U_2-c\right)}}+\frac{1}{3} e^{\frac{3 p_1}{\sigma\left(U_2-c\right)}}-\frac{1}{3}\right)}\cdot\\
& \left\{\operatorname { l n } \left( \frac { \rho _ { 1 } } { \rho _ { 3 } } \right) \left[-2 \rho_1\left(-U_2+c\right) p_1 p_3 \rho_3 e^{\frac{p_1}{2 \sigma\left(U_2-c\right)}}-2 \rho_1\left(-U_2+c\right) p_1 p_3 \rho_3 e^{\frac{5 p_1}{2 \sigma\left(U_2-c\right)}}\right.\right. \\
& -\frac{2}{3}\left(\rho_1^2+\rho_3^2\right)\left(-U_2+c\right) p_1 p_3 e^{\frac{5 p_1}{2 \sigma\left(U_2-c\right)}}+\rho_3 \sqrt{\frac{\rho_1}{\rho_3}}\left(\left(p_3\left(\rho_1+\rho_3\right)\left(-U_2+c\right) p_1\right.\right. \\
& -\sigma^2\left(U_2-c\right)^4\left(\rho_1-\rho_3\right) \ln ^2\left(\frac{\rho_1}{\rho_3}\right)+6 \sigma\left(\rho_1-\rho_3\right) d\left(U_2-c\right)^2 \ln \left(\frac{\rho_1}{\rho_3}\right) \\
& \left.+6 d^2\left(\rho_1-\rho_3\right)\right) e^{\frac{2 p_1}{\sigma\left(U_2-c\right)}}+\frac{1}{3}\left(p_1 p_3\left(\rho_1+\rho_3\right)\left(-U_2+c\right)+p_5\left(\rho_1-\rho_3\right)\right) e^{\frac{3 p_1}{\sigma\left(U_2-c\right)}} \\
& +\left(p_1 p_3\left(\rho_1+\rho_3\right)\left(-U_2+c\right)+\sigma^2\left(-U_2+c\right)^4\left(\rho_1-\rho_3\right) \ln ^2\left(\frac{\rho_1}{\rho_3}\right)\right. \\
& \left.-6 \sigma\left(\rho_1-\rho_3\right) d\left(-U_2+c\right)^2 \ln \left(\frac{\rho_1}{\rho_3}\right)+6 d^2\left(\rho_1-\rho_3\right)\right) e^{\frac{p_1}{\sigma\left(U_2-c\right)}} \\
& \left.\left.\left.+\frac{1}{3} p_1 p_3\left(\rho_1+\rho_3\right)\left(-U_2+c\right)-\frac{1}{3} p_5\left(\rho_1+\rho_3\right)\right)\right]\right\}, \\
&
\end{split}&
\end{flalign}

\begin{flalign}
\begin{split}\label{100}
& 2 I \cdot a_3=\frac{1}{\sqrt{p_3 \ln \left(\frac{\rho_1}{\rho_3}\right) \sigma} d\left(-p^{2}_1+\sigma^{2} \ln^{2} \left(\frac{\rho_1}{\rho_3}\right)\left(-U_2+c\right)^{2}\right) \sigma}\cdot \\
& \frac{1}{\left(e^{\frac{p_1 d+\sigma \ln \left(\frac{\rho_1}{\rho_3}\right)(2 D+d)\left(-U_2+c\right)}{2 d \sigma\left(-U_2+c\right)}}-e^{\frac{-p_1 d+\sigma \ln \left(\frac{\rho_1}{\rho}\right)(2 D+d)\left(-U_2+c\right)}{2 d \sigma\left(-U_2+c\right)}}\right)^2}\cdot \\
& \left\{2 \rho _ { 3 }\ln ^3\left(\frac{\rho_1}{\rho_3}\right) \left[\left(4 p_1 d+4 p_2 \sigma\left(-U_2+c\right)\right) e^{\frac{-p_1 d+\sigma \ln \left(\frac{\rho_1}{\rho_3}\right)(4 D+3 d)\left(-U_2+c\right)}{2 d \sigma\left(-U_2+c\right)}}\right.\right. \\
& +\left(4 p_1 d-4 p_2 \sigma\left(-U_2+c\right)\right) e^{\frac{p_1 d+\sigma \ln \left(\frac{\rho_1}{\rho_3}\right)(4 D+3 d)\left(-U_2+c\right)}{2 d \sigma\left(-U_2+c\right)}} \\
& -\left(\left(U_2-c\right) p_1+p_{2}\right)\left(-U_2+c\right) \sigma e^{\frac{-p_1 d+\sigma \ln \left(\frac{\rho_1}{\rho_3}\right)(2 D+d)\left(-U_2+c\right)}{2 d \sigma\left(-U_2+c\right)}} \\
& -\left(\left(-U_2+c\right) p_1+p_{2}\right)\left(-U_2+c\right) \sigma e^{\frac{-p_1 d+\sigma \ln \left(\frac{\rho_1}{\rho_3}\right)(D+d)\left(-U_2+c\right)}{2 d \sigma\left(-U_2+c\right)}} \\
& +\left(\left(-U_2+c\right) p_1+p_{2}\right)\left(-U_2+c\right) \sigma e^{\frac{p_1 d+\sigma \ln \left(\frac{\rho_1}{\rho_3}\right)(2 D+d)\left(-U_2+c\right)}{2 d \sigma\left(-U_2+c\right)}} \\
& +\left(\left(U_2-c\right) p_1+p_{2}\right)\left(-U_2+c\right) \sigma e^{\frac{p_1 d+\sigma \ln \left(\frac{\rho_1}{\rho_3}\right)(D+d)\left(-U_2+c\right)}{2 d \sigma\left(-U_2+c\right)}} \\
& \left.\left.+2\left(\left(\sigma\left(U_2-c\right)^2-2 d\right)\left(\frac{\rho_1}{\rho_3}\right)^{\frac{2 D+2 d}{d}}-\left(\sigma\left(U_2-c\right)^2+2 d\right)\left(\frac{\rho_1}{\rho_3}\right)^{\frac{2 D+d}{d}}\right) p_1\right] \right\}, \\
&
\end{split}&
\end{flalign}

\begin{flalign}
\begin{split}\label{99}
& 2 I \cdot a_2=\frac{1}{3} \rho_1\left(-U_1+c\right)^2 D+\frac{1}{3\left(-\sigma\left(-U_3+c\right)^2 H-D-d\right)^2}\cdot \\
&  \rho_3 \left(-U_3+c\right)^2(H-D-d)\left(3 \sigma^2\left(-U_3+c\right)^4\right. \\
& \left.-3 \sigma\left(-U_3+c\right)^2(H-D-d)+(H-D-d)^2\right) \\
& \frac{1}{\sqrt{\frac{\rho_1}{\rho_3}} \ln \left(\frac{\rho_1}{\rho_3}\right) \sigma p_3\left(-e^{\frac{2 p_1}{2 \sigma\left(U_2-c\right)}}+2 e^{\frac{p_1}{2 \sigma\left(U_2-c\right)}}-1\right)}\cdot \\
& \left\{( - U _ { 2 } + c ) ^ { 2 } d \left(4 p_1\left(-\frac{1}{2} p_1+\sigma\left(-U_2+c\right)\right) e^{\frac{p_1}{2 \sigma\left(U_2-c\right)}}\right.\right.\\
&  -4 p_1\left(\frac{1}{2} p_1+\sigma\left(-U_2+c\right)\right) e^{\frac{3 p_1}{2 \sigma\left(-U_2+c\right)}} \\
& \left.\left.+\left(\sigma\left(-U_2+c\right) e^{\frac{2 p_1}{\sigma\left(U_2-c\right)}}+2 p_1 e^{\frac{p_1}{\left(U_2-c\right)}}-\sigma\left(-U_2+c\right)\right)\left(\rho_1+\rho_3\right) \sqrt{\frac{\rho_1}{\rho_3}}\right)\right\},
\end{split}&
\end{flalign}
where
\begin{flalign}
\begin{split}\label{101}
&p_{1}=\sqrt{\sigma\left(\sigma(-U_{2}+c)^2 \ln \left(\frac{\rho_{ 1}}{\rho_{ 3}}\right)-4 d\right) \ln \left(\frac{\rho_{1}}{\rho_{3}}\right)},\\
&p_{2}=\sigma(-U_{2}+c)^2  \ln \left(\frac{\rho_{ 1}}{\rho_{ 3}}\right)-2 d,\\
&p_{3}=\sigma(-U_{2}+c)^2  \ln \left(\frac{\rho_{ 1}}{\rho_{ 3}}\right)-4 d,\\
&p_{4}=\sigma(-U_{2}+c)^2  \ln \left(\frac{\rho_{ 1}}{\rho_{ 3}}\right)-\frac{9 d}{2},\\
&p_{5}=\sigma^{2}(-U_{2}+c)^4\ln^{2} \left(\frac{\rho_{ 1}}{\rho_{ 3}}\right)-6\sigma d(-U_{2}+c)^2 \ln \left(\frac{\rho_{ 1}}{\rho_{ 3}}\right)+6d^{2}.\\
\end{split}&
\end{flalign}

\end{document}